\documentclass[a4paper,11pt]{article}
\pdfoutput=1
\usepackage[utf8]{inputenc}
\usepackage[T1]{fontenc}
\usepackage{jheppub}
\usepackage{natbib}

\usepackage[usenames,dvipsnames,svgnames,table,x11names]{xcolor}
\usepackage{lmodern,amsmath,multirow,cancel}
\usepackage{booktabs}
\usepackage{xstring}
\usepackage{ifthen}
\usepackage{bbm}

\usepackage{ascmac,braket,bm,mathrsfs,amsthm,amsfonts}
\usepackage{hyperref}
\usepackage{graphicx}
\usepackage{subcaption} % Allow for subfigures
\usepackage{comment}
\usepackage[normalem]{ulem}

\usepackage{titlesec}
\newcommand*{\justifyheading}{\raggedright}
\titleformat{\chapter}[display]
  {\normalfont\huge\bfseries\justifyheading}{\chaptertitlename\ \thechapter}
  {20pt}{\Huge}
\titleformat{\section}
  {\normalfont\Large\bfseries\justifyheading}{\thesection}{1em}{}
\titleformat{\subsection}
  {\normalfont\large\bfseries\justifyheading}{\thesubsection}{1em}{}
\titleformat{\subsubsection}
  {\normalfont\bfseries\justifyheading}{\thesubsubsection}{1em}{}
\usepackage[english]{babel}
\usepackage[dvipsnames]{xcolor}
\numberwithin{equation}{section}

\usepackage{slashed}
\newcommand{\Slash}[1]{{\ooalign{\hfil/\hfil\crcr$#1$}}}

\newcommand{\eg}{{\em e.g.}}
\newcommand{\ie}{{\em i.e.}}

\newcommand{\non}{\nonumber \\}

\usepackage{adjustbox}
\usepackage{tabularx}
\newcolumntype{Y}{>{\centering\arraybackslash}X} %for tabularx
\usepackage{booktabs} %for \toprule
\usepackage{colortbl} %for \rowcolors
\def\beq#1\eeq{\begin{align}#1\end{align}}
\RequirePackage{xspace}
\def\Bbar    {\kern 0.18em\overline{\kern -0.18em B}{}\xspace}

\def\Kbar    {\kern 0.18em\overline{\kern -0.18em K}{}\xspace}
\def\Kb      {\ensuremath{\Kbar}\xspace}

\renewcommand*{\backref}[1]{}
\renewcommand*{\backrefalt}[4]{%
  \ifcase #1 %
    \relax % use \relax if you do not want the "No citations" message
  \or
    {\scriptsize (page~#2).}%
  \else
    {\scriptsize (pages~#2).}%
  \fi%
}

\definecolor{light-blue}{rgb}{0.3, 0.3, 1}
\definecolor{kit-green}{rgb}{0, 
0.58823 %150/255
, 0.50980 %130/255
}

\preprint{CHIBA-EP-275}

\title{Complete one-loop QED corrections to \texorpdfstring{\boldmath{$D_s^+$}}{Ds+} leptonic decays and impact on the CKM unitarity test} 

\author[a,b]{Teppei Kitahara,}
\author[c,a]{Jun Miyamoto,}
\author[d,a]{and Kota Sasaki}

\affiliation[a]{
Department of Physics, Graduate School of Science,
Chiba University, Chiba 263-8522, Japan}
\affiliation[b]{
  Kobayashi-Maskawa Institute for the Origin of Particles and the
  Universe, Nagoya University,
  Furo-cho Chikusa-ku, Nagoya 464-8602 Japan
}
\affiliation[c]{
ICRR, The University of Tokyo, Kashiwa, Chiba 277-8582, Japan}
\affiliation[d]{
Department of Physics, Graduate School of Science and Engineering, Chiba University, 263-8522, Chiba, Japan}

\emailAdd{kitahara@chiba-u.jp}
\emailAdd{m38j@icrr.u-tokyo.ac.jp}
\emailAdd{sasaki@chiba-u.jp}

\abstract{
Recently, a violation of the CKM unitarity condition has been reported in the latest charm-meson data with the latest lattice results, once the universal electroweak correction is taken into account.
In this article, 
we analytically derive for the first time the complete one-loop electroweak (EW) and QED corrections to the $D_{s}^+ \to \ell^+ \nu_\ell$ decays for $\ell = \mu, \tau$.
Our analysis incorporates both short-distance EW-QED corrections, which are beyond the leading-logarithmic approximation (the so-called Sirlin factor), 
and long-distance soft-photon corrections that depend on the maximum total energy of undetected photons with their resummation.
Although the inclusive QED corrections to the meson leptonic decays are well known,
they do not match the actual measurement circumstances in $D_s^+ \to \mu^+ \nu_\mu$.
We find $
|V_{cs}|_{D_s} = 0.991 \pm 0.007 $
 from the latest data on $D_s^+$ leptonic decays.
We show that properly 
including these radiative corrections is essential to bring the second-column CKM unitarity tests into agreement with the Standard Model expectation.
The study emphasizes that the current limiting factor in confirming CKM unitarity is the precision of QED corrections, and it points out that improving lattice simulations, taking the QED corrections into account, would be desirable for a more robust confirmation.
}


\begin{document}
\sloppy %https://tex.stackexchange.com/questions/9107/how-can-i-make-my-text-never-go-over-the-right-margin-by-always-hyphenating-or-b

% get rid of JHEP header
\makeatletter\renewcommand{\@fpheader}{\ }\makeatother

\maketitle

\renewcommand{\thefootnote}{\#\arabic{footnote}}
\setcounter{footnote}{0}

%=======================================================
%        memo
%=======================================================
%\section*{memo}
%\begin{itemize}
%\end{itemize}

%=======================================================
%        INTRODUCTION
%=======================================================
\section{Introduction}
The Cabibbo–Kobayashi–Maskawa (CKM) matrix, $V$, expresses the mixing among quark flavors and is also known as the only source of (confirmed) $CP$ violation \cite{Kobayashi:1973fv}. 
Within the Standard Model (SM), 
the unitarity of the CKM matrix
\beq
VV^\dag = V^\dag V=\mathbbm{1}\,,
\eeq 
is imposed without any question.
This unitarity condition arises from the Higgs mechanism, which involves only the Yukawa coupling, generation-independent gauge interactions, the left-handedness of the charged current, and the absence of exotic matter in the full Lagrangian. 
On the other hand, it is also well known that the SM is not the theory of everything because of, \eg, the matter-antimatter asymmetry in the Universe, and it is expected that New Physics (NP) beyond the SM will emerge at some high energy scale.
For instance, in the vector-like quark \cite{Belfatto:2019swo,Cheung:2020vqm,Belfatto:2021jhf,Branco:2021vhs,Crivellin:2022rhw} or the vector-like lepton models \cite{Endo:2020tkb,Crivellin:2020ebi,Kirk:2020wdk}, the unitarity of the $3\times 3$ CKM matrix will be violated  to some extent.

There are two categories of the unitarity conditions.
The first one is the orthogonality conditions:
\beq
\sum_{k=1}^3 V_{ik} V^\ast_{jk} = \sum_{k=1}^3 V_{ki} V^\ast_{kj}= 0 \quad (\text{for~} i\neq j)\,, 
\eeq 
which is also called the unitarity triangles. 
The advantage is that it allows us to verify the consistency of magnitudes of CP‑violating observables across different mesons.
In particular, the condition of $V_{ud}V^\ast_{ub} + V_{cd}V^\ast_{cb} + V_{td}V^\ast_{tb} =0$ has been comprehensively studied by the CKMfitter \cite{Hocker:2001xe,Charles:2004jd} and UTfit collaborations \cite{UTfit:2005ras,UTfit:2007eik}, and no indication of NP has been observed, up to now.\footnote{%
Furthermore, the kaon unitarity triangle has also been studied in Refs.~\cite{Buchalla:1996fp,Buras:2006gb,Lehner:2015jga,Lunghi:2024sjy,Dery:2025pcx}.}
On the other hand, the second category is the normalization conditions:
\beq
\sum_{k=1}^3 \left|V_{ik} \right|^2 = \sum_{k=1}^3 \left|V_{ki} \right|^2 = 1\,.
\label{eq:normalization}
\eeq 
Historically, studies of these conditions have been limited compared to the unitarity triangles because the theoretical uncertainties from the form factors have covered the underlying trend, rendering detailed discussion challenging.  
However, recent progress in lattice QCD simulations significantly reduced these theoretical uncertainties and enabled quantitative analyses of the normalization conditions \cite{FlavourLatticeAveragingGroupFLAG:2024oxs}.

In recent years, a notable tension has become clear in the first-row CKM unitarity test: $|V_{ud}|^2 + |V_{us}|^2 + |V_{ub}|^2 - 1 = -(1.51 \pm 0.53) \times 10^{-3}$, where a slight deficit was reported \cite{Crivellin:2022rhw}.
The corresponding significance is $-2.8\sigma$.\footnote{%
This significance increases (about $+1\sigma$)
when the neutron lifetime data based on the neutron-beam experiments \cite{Yue:2013qrc} are adopted in the global fit analysis \cite{Kitahara:2023xab}.
On the other hand, this tension slightly decreases (about $-0.5\sigma$) when a new lattice calculation for the QED correction to the beta decays \cite{Ma:2023kfr} is used.
More recent progress on the QED corrections and the finite nuclear size corrections to the beta decays are provided in Refs.~\cite{Moretti:2025qxt,Cao:2025zxs,Crosas:2025xyv} and Refs.~\cite{Seng:2022epj,Seng:2023cgl,Gorchtein:2025wli}, respectively.} 
This discrepancy has been referred to as the Cabibbo angle anomaly (CAA) \cite{Grossman:2019bzp,Coutinho:2019aiy}.
The test of the CAA has stimulated intense surveys of the QED radiative corrections in the beta decays and kaon decays, including the lattice QCD$+$QED simulations \cite{Carrasco:2015xwa,Lubicz:2016xro,Giusti:2017dwk,DiCarlo:2019thl,Frezzotti:2020bfa}. 
It should be equally important to extend such unitarity tests to other rows and columns of the CKM matrix.

According to the approach of the Particle Data Group (PDG),
the CKM elements $|V_{cd}|$ and $|V_{cs}|$ are extracted from the $D$ and $D_s^+$ meson leptonic and semi-leptonic decays, using the theoretical knowledge of the form factors. 
The PDG approach is based on the tree-level matching of the theoretical predictions 
with the charm-meson measurements 
because the radiative corrections have not been fully studied.
However, it was recently pointed out that once the universal electroweak correction, corresponding to a short-distance correction, is taken into account in the extraction of $|V_{cs}|$, 
the normalization conditions in Eq.~\eqref{eq:normalization} are significantly violated at $-4.3\sigma$ level 
for the second-row unitarity condition and at $-5.2\sigma$ level for the second-column one
\cite{Bolognani:2024cmr}.

It is important to note that the Wolfenstein parametrization for the CKM matrix \cite{Wolfenstein:1983yz} follows strict unitarity to order of $\lambda^3$, while the magnitude of the soft photon correction, corresponding to a long-distance correction, is typically
\beq
\mathcal{O}\left( \alpha\right) \approx \mathcal{O}\left(\lambda^3\right)\,.
\eeq
It can occasionally provide nontrivial and sizable corrections exceeding $\mathcal{O}(1)\%$, \eg, the long-distance QED corrections to the lepton flavor universality test \cite{deBoer:2018ipi}.
Therefore, such $\mathcal{O}\left( \alpha\right)$ corrections can easily ruin the unitarity condition within the Wolfenstein parametrization.
In other words, complete one-loop QED corrections are needed in the extractions of the CKM components to check the unitarity condition.

In actual experiments, we will explain later that there is a cut that excludes a hard photon emission in order to collect signals in  $D_s^+ \to \mu^+ \nu_\mu$ process. 
On the other hand, such a cut does not work in  $D_s^+ \to \tau^+ \nu_\tau$ process. 
This difference would provide a nontrivial effect for the extraction of $|V_{cs}|$.

In this article,
we  analytically study the complete one-loop radiative corrections to $D_s^+ \to \ell^+ \nu_\ell$ decays.
In Sec.~\ref{sec:SD}, we  derive the one-loop analytic formula of the short-distance EW-QED corrections,
reproducing the known leading-logarithmic correction, and also obtain a non-logarithmic correction. Furthermore,
we discuss the regularization-scheme dependence.
In Sec.~\ref{sec:long},
we  compute the long-distance QED corrections and provide the analytic formulae.
In Sec.~\ref{sec:calc}, 
we  first estimate the maximum energy of the undetected soft photon in the experimental setup for $D_s^+ \to \ell^+ \nu_\ell$ measurements. Then, we  discuss the PHOTOS corrections in the experimental results. Based on these evaluations,
we extract the value of $|V_{cs}|$ from the latest data, and investigate the CKM unitarity test. 
Section~\ref{sec:conc} is
devoted to the conclusions and discussion.
Various useful results and further detailed calculations are given in the appendices.

%=======================================================
%        SHORT DISTANCE
%=======================================================
\section{Short-distance QED corrections}
\label{sec:SD}

In this section, we calculate the short-distance EW-QED corrections to the $c s \ell \nu $ amplitude. 
In order to consistently treat the Fermi coupling constant $G_F$,  
 the short-distance EW-QED corrections to the muon lifetime are also discussed.
Throughout this paper, we use the QED charge defined below:
\beq
Q_c = \frac{2}{3}\,, \quad Q_s = - \frac{1}{3}\,, \quad  Q_\ell = -1\,, \quad Q_\nu =0\,.
\eeq

Note that thanks to the QCD Ward identity, QCD one-loop corrections are completely canceled among the self-energies and the vertex correction in the $cs\ell \nu$ amplitude, and 
there is no operator mixing with the other operators in the weak Hamiltonian \cite{Buras:1998raa}.
The leading QCD correction appears from two-loop level as a mixed QCD-QED correction \cite{Sirlin:1977sv,Sirlin:1981ie,Marciano:1983ss,Marciano:1985pd}. 
We will see the impact at the end of this section.

\subsection{Corrections to the \texorpdfstring{$cs\ell\nu$}{cslnu} amplitude}

\paragraph{Electroweak corrections.}
\begin{figure}[t]
    \centering
    \begin{subfigure}{.3 \textwidth}
        \centering
        \includegraphics[width=1.\linewidth]{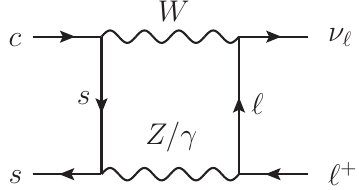}
        \caption{}
    \end{subfigure}
    \qquad 
    \begin{subfigure}{.3 \textwidth}
        \centering
        \includegraphics[width=1.\linewidth]{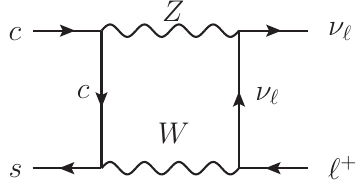}
        \caption{}
    \end{subfigure}  \\[0.5cm]
    \begin{subfigure}{.3 \textwidth}
        \centering
        \includegraphics[width=1.\linewidth]{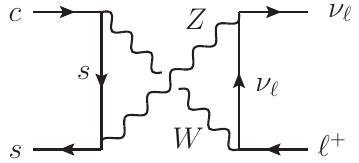}
        \caption{}
    \end{subfigure}
    \qquad 
    \begin{subfigure}{.3 \textwidth}
        \centering
        \includegraphics[width=1.\linewidth]{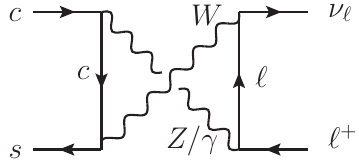}
        \caption{}
    \end{subfigure}        
        \caption{Feynman diagrams for the box contributions.
        }
        \label{fig:box}
\end{figure}
First, we calculate the $ZW$-box contributions to the $c s \ell \nu $ amplitude.
There are four one-loop diagrams shown in Fig.~\ref{fig:box}, all of which give finite contributions, which therefore are regularization scheme independent. 
The results for individual diagrams are
\beq
i\mathcal{M}_{\text{(a)}}^{ZW} &= i\mathcal{M}_0\frac{g_Z^2}{16\pi^2}\left(-\frac{1}{2}-Q_ss_W^2\right)\left(-\frac{1}{2}-Q_\ell s_W^2\right)\frac{M_W^2}{M_Z^2-M_W^2}\ln{\frac{M_Z^2}{M_W^2}}+ \mathcal{O}\left(\frac{s}{M_W^2}\right)\,,\\
%%%
i\mathcal{M}_{\text{(b)}}^{ZW} & = i\mathcal{M}_0\frac{g_Z^2}{32\pi^2}\left(\frac{1}{2}-Q_cs_W^2\right)\frac{M_W^2}{M_Z^2-M_W^2}\ln{\frac{M_Z^2}{M_W^2}}+ \mathcal{O}\left(\frac{s}{M_W^2}\right)
\,,\\
%%%
i\mathcal{M}_{\text{(c)}}^{ZW}& = - i\mathcal{M}_0\frac{g_Z^2}{8\pi^2}\left(-\frac{1}{2}-Q_ss_W^2\right)\frac{M_W^2}{M_Z^2-M_W^2}\ln{\frac{M_Z^2}{M_W^2}}+ \mathcal{O}\left(\frac{s}{M_W^2}\right)\,,\\
%%%
i\mathcal{M}_{\text{(d)}}^{ZW} & = - i\mathcal{M}_0\frac{g_Z^2}{4\pi^2}\left(\frac{1}{2}-Q_cs_W^2\right)\left(-\frac{1}{2}-Q_\ell s_W^2\right)\frac{M_W^2}{M_Z^2-M_W^2}\ln{\frac{M_Z^2}{M_W^2}}
+ \mathcal{O}\left(\frac{s}{M_W^2}\right)\,,
%%%%
\eeq
where the tree-level amplitude is defined by 
\beq
\mathcal{M}_0 = - \frac{4 G_F}{\sqrt{2} } V_{cs}^\ast (\bar{s} \gamma^{\mu} P_L c ) (\bar\nu_\ell \gamma_\mu P_L \ell )\,,
\label{eq:tree0}
\eeq
with $G_F/\sqrt{2} \equiv g^2/{8 M_W^2}$,  
$g_Z \equiv \sqrt{g^2 + g^{\prime 2}}$, and the weak-mixing angle in the $\overline{\text{MS}}$ scheme 
$s_W^2 \equiv \sin^2 \hat{\theta}_W (M_Z) = 0.23129(4)$ \cite{ParticleDataGroup:2024cfk}. 
In these calculations, we kept $s = (p_c + p_{{s}})^2$ terms in the loop integrals, and it was found that all of them are suppressed by $s/M_W^2$; 
When meson decays are considered with $\sqrt{s} = \mathcal{O}(1)\,$GeV, these terms are safely negligible in the radiative corrections. 
Here and throughout, we used the Feynman-'t~Hooft gauge %($\xi=1$) 
for the gauge fixing of the weak gauge boson propagators. 
Since all the radiative corrections in this study are flavor conserving, 
the Nambu–Goldstone (NG) boson contributions in the loop diagrams are always proportional to small Yukawa couplings. 
So, these contributions are additionally suppressed by $m_{\rm light}^2/M_W^2$ and are also safely negligible.

The total contribution from the above $ZW$-box diagrams is
\beq
i\mathcal{M}^{ZW} & \simeq  
i\mathcal{M}_0\frac{g_Z^2}{32\pi^2}
\left[5  + 5 \left(-Q_c + Q_s + Q_\ell \right) s_W^2 + 2 Q_\ell \left( - 4 Q_c +  Q_s \right)s_W^4\right]
\frac{M_W^2}{M_Z^2-M_W^2}\ln{\frac{M_Z^2}{M_W^2}}\nonumber \\
 & = i\mathcal{M}_0\frac{\alpha(M_Z)}{4\pi} 
  \left[Q_\ell \left( - 4 Q_c +  Q_s \right) + \frac{5}{2s_W^2}  \left(\frac{1}{s_W^2}  -Q_c + Q_s + Q_\ell  \right)\right]
\ln{\frac{M_Z^2}{M_W^2}}
\,.
\label{eq:SDZWbox}
\eeq
Here, $4 \pi \alpha (M_Z) = g^2 s^2_W $ is the (squared) electromagnetic coupling in the $\overline{\text{MS}}$ scheme and $\alpha(M_Z)^{-1} =  127.930(8) $ \cite{Erler:1998sy,ParticleDataGroup:2024cfk}.

One should note that in traditional calculations, the Fermi coupling constant is related to the renormalized $SU(2)_L$ gauge coupling and $W$-boson mass as 
\beq
\frac{G_F}{\sqrt{2}}=\frac{g^2}{8 M_W^2}\left(1+\Delta r\right)\,,
\eeq
where $\Delta r$ stands for the short-distance radiative corrections to the muon lifetime  
\cite{Sirlin:1980nh,Stuart:1991cc,vanRitbergen:1999fi}, which were evaluated using the naive dimensional regularization (NDR) \cite{Buras:1989xd}.
For the clarity,  
we do not use this form.
Instead, we explicitly compute 
the short-distance corrections to the muon lifetime in this section and subtract them from our results, whose prescription is equivalent to
$(G_F/\sqrt{2}) (1 - \Delta r)=g^2/8 M_W^2$.
This also allows us to discuss the dependence on the regularization schemes.

\paragraph{$\gamma W$-box contributions.}
Next, we consider the $\gamma W$-box contributions. 
The one-loop diagrams in the full theory are shown in Fig.~\ref{fig:box}(a) and (d).
To regulate the IR singularities coming from the massless photon propagator, 
here and throughout, we will introduce a tiny nonzero photon mass $m_\gamma$. 
By explicit loop calculations in the full theory,
we obtain: 
\beq
%\begin{aligned}
i \mathcal{M}_{\text{(a)}}^{\gamma W}
&= 
- i \mathcal{M}_0 \frac{\alpha}{ 4 \pi} Q_s Q_{\ell}\left\{\frac{7}{2} -\ln \left[\frac{2 M_W^2}{s(1-\cos \theta)}\right]+\frac{2\pi^2}{3}+\ln ^2\left[\frac{2 m_\gamma^2}{s(1-\cos \theta)}\right] \right. \non
& \qquad \left. + 4 \ln \left[\frac{2 m_\gamma^2}{s(1-\cos \theta)}\right]\right\} +  \mathcal{O}\left(\frac{s}{M_W^2}\right) \,,\\
%\end{aligned}\\
%%%%%
%\begin{aligned}
i \mathcal{M}_{\text{(d)}}^{\gamma W}
&= 
i \mathcal{M}_0 \frac{\alpha}{ \pi} Q_c Q_{\ell}\left\{-\ln \left[\frac{2 M_W^2}{s(1+\cos \theta)}\right]+\frac{\pi^2}{6}+\frac{1}{4} \ln ^2\left[\frac{2 m_\gamma^2}{s(1+\cos \theta)}\right]
 \right. \non
& \qquad \left. 
+\ln \left[\frac{2 m_\gamma^2}{s(1+\cos \theta)}\right]\right\} +  \mathcal{O}\left(\frac{s}{M_W^2}\right) \,,
%\end{aligned}
%%%%%
\eeq
where $\theta$ is the angle between the momenta of $c$ and $\nu_\ell$.

\begin{figure}[t]
    \centering
    \begin{subfigure}{.3 \textwidth}
        \centering
        \includegraphics[width=1.\linewidth]{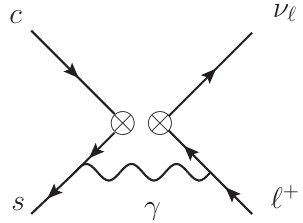}
        \caption{}
    \end{subfigure}
    \qquad 
    \begin{subfigure}{.3 \textwidth}
        \centering
        \includegraphics[width=1.\linewidth]{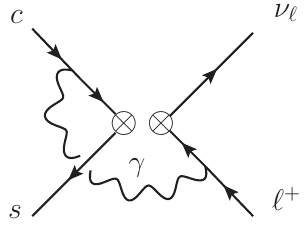}
        \caption{}
    \end{subfigure}    
        \caption{Feynman diagrams for the one-loop matchings onto the $c s \ell \nu$ operator.
        The vertex         ``$\otimes~\otimes$'' denotes the insertion of the four-fermion operator $(\bar{s}\gamma^{\mu}P_L c)(\bar{\nu}_\ell \gamma_\mu P_L \ell)$. 
        }
        \label{fig:EFT-one-loop}
\end{figure}

For the one-loop matching onto the weak Hamiltonian at the renormalization scale $\mu \ll M_W $, we need the calculations of the one-loop corrections to the four-fermion operator $(\bar{s}\gamma^{\mu}P_L c)(\bar{\nu}_\ell \gamma_\mu P_L \ell)$
shown in Fig.~\ref{fig:EFT-one-loop}.
These diagrams contain both UV and IR divergences.
To assess the UV divergence, we considered two different schemes of the dimensional regularization ($d = 4 - 2 \epsilon$), although  
depending on the regularization schemes the finite terms in the matching conditions are changed. 
For the one-loop corrections to the operator, we obtain
\beq
i \mathcal{M}_{\text{(a)}}^{\gamma}& =- i \mathcal{M}_0 \frac{\alpha}{4 \pi} Q_s Q_\ell \left\{-\frac{1}{\bar{\epsilon}} +4 +\kappa_a \right. \non
& \qquad \left. - \ln \left[\frac{2 \mu^2}{s(1-\cos \theta)}\right] +\frac{2\pi^2}{3}+ \ln ^2\left[\frac{2 m_\gamma^2}{s(1-\cos \theta)}\right]+ 4 \ln \left[\frac{2 m_\gamma^2}{s(1-\cos \theta)}\right]\right\}\,,\\
%%%%%%
i \mathcal{M}_{\text{(b)}}^{\gamma}& =i \mathcal{M}_0 \frac{\alpha}{\pi} Q_c Q_\ell  \left\{-\frac{1}{\bar{\epsilon}}-\frac{5}{4}
+\kappa_b \right. \non
& \qquad \left. -\ln \left[\frac{2 \mu^2}{s(1+\cos \theta)}\right]+\frac{\pi^2}{6}+\frac{1}{4} \ln ^2\left[\frac{2 m_\gamma^2}{s(1+\cos \theta)}\right]+\ln \left[\frac{2 m_\gamma^2}{s(1+\cos \theta)}\right]\right\}\,,
\eeq
where $1/\bar{\epsilon} \equiv 1/\epsilon - \gamma_{\rm E} + \ln 4 \pi$ and $\gamma_{\rm E} $ is the Euler’s constant. 
The UV divergences ($1/\bar{\epsilon}$) are absorbed by the operator renormalization. 
The parameters $\kappa_i$ represent the regularization-scheme-dependent parameters
that arise from the treatment of $\gamma_5$ in $d$ dimensions. 
We parameterized them as 
$\kappa_i =0$ in the 
NDR scheme  
with anticommuting $\gamma_5$.
On the other hand, the additional constants 
$\kappa_a = -2, \kappa_b =0 $ appear 
in the 't~Hooft--Veltman (HV) scheme with non-anticommuting $\gamma_5$ \cite{tHooft:1972tcz, Bollini:1972ui, Breitenlohner:1977hr}. See Appendix~\ref{sec:Appreg}
 for the scheme dependence.

Compared with these formulae, 
one can obtain proper one-loop level radiative corrections at the renormalization scale $\mu$, as follows:
\beq
%%%%%%
i \mathcal{M}_{\text{(a)}}^{\gamma W} -  \left. i \mathcal{M}_{\text{(a)}}^{\gamma }\right|_{\bar{\epsilon}^0}
&=  i \mathcal{M}_0 \frac{\alpha}{ 4\pi} Q_s Q_\ell\left[  \ln \left(\frac{ M_W^2}{\mu^2 }\right)+\frac{1}{2} + \kappa_a  \right]\,,
\label{eq:cancel1}\\
%%%%%
i \mathcal{M}_{\text{(d)}}^{\gamma W} -  \left. i \mathcal{M}_{\text{(b)}}^{\gamma}\right|_{\bar{\epsilon}^0}
&= - i \mathcal{M}_0 \frac{\alpha}{ \pi} Q_c Q_\ell\left[  \ln \left(\frac{ M_W^2}{\mu^2 }\right)- \frac{5}{4}  + \kappa_b \right]\,.
\label{eq:cancel2}
\eeq
Here, the $\theta$ and $m_\gamma$ divergences are thoroughly canceled out as expected.
For the details of the loop calculations, see 
Appendix~\ref{sec:WET}.
The total contribution from the $\gamma W$-box diagrams is
\beq
i \mathcal{M}^{\gamma W} 
&= i \mathcal{M}_0 \frac{\alpha}{ 4\pi} 
\left[   Q_\ell \left( - 4 Q_c + Q_s \right) \ln \left(\frac{ M_W^2}{\mu^2 }\right)
+  Q_\ell \left( 5 Q_c + \frac{1}{2}Q_s\right)
+ Q_\ell \left( Q_s\kappa_a - 4 Q_c \kappa_b \right)\right]\,.
\eeq

\paragraph{Vertex corrections.}

\begin{figure}[t]
    \centering
    \begin{subfigure}{.32 \textwidth}
        \centering
        \includegraphics[width=1.\linewidth]{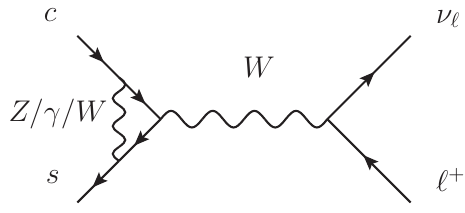}
        \caption{}
    \end{subfigure}
    \qquad 
    \begin{subfigure}{.3 \textwidth}
        \centering
        \includegraphics[width=1.\linewidth]{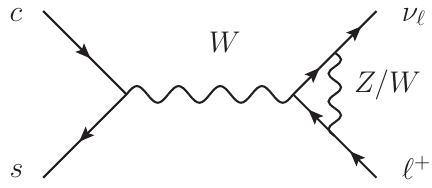}
        \caption{}
    \end{subfigure}         
        \caption{Vertex corrections with virtual gauge bosons attached between fermions. 
        }
    \label{fig:vertex-1}
\end{figure}
\begin{figure}[t]
    \centering
    \begin{subfigure}{.3 \textwidth}
        \centering
        \includegraphics[width=1.\linewidth]{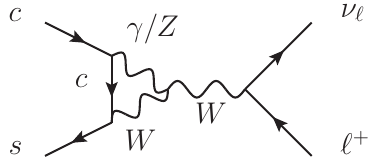}
        \caption{}
    \end{subfigure}
    \qquad 
    \begin{subfigure}{.3 \textwidth}
        \centering
        \includegraphics[width=1.\linewidth]{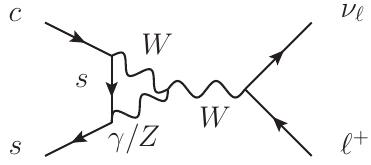}
        \caption{}
    \end{subfigure}  \\[0.5cm]
    \begin{subfigure}{.3 \textwidth}
        \centering
        \includegraphics[width=1.\linewidth]{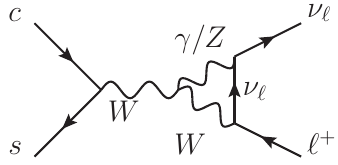}
        \caption{}
    \end{subfigure}
    \qquad 
    \begin{subfigure}{.3 \textwidth}
        \centering
        \includegraphics[width=1.\linewidth]{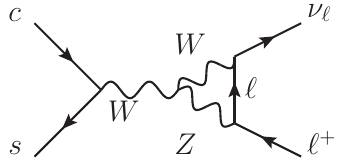}
        \caption{}
    \end{subfigure}        
        \caption{Vertex corrections involving the three-gauge-boson interactions. 
        }
        \label{fig:vertex-2}
\end{figure}

There are two types of vertex corrections;
the vertex corrections shown in Fig.~\ref{fig:vertex-1}
that the virtual gauge bosons couple to fermions, and diagrams shown in Fig.~\ref{fig:vertex-2} that involve the three-gauge-boson interactions.
It is well known that these contributions are completely canceled in total.
Let us present this briefly for pedagogical purposes.

First, the photon correction in Fig.~\ref{fig:vertex-1}(a) is completely canceled (with the corresponding one in Fig.~\ref{fig:EFT-one-loop}) in the one-loop matching onto the weak Hamiltonian, \eg, see Ref.~\cite{Buras:1998raa}.

\begin{figure}[t]
    \centering
    \begin{subfigure}{.26 \textwidth}
        \centering
        \includegraphics[width=1.\linewidth]{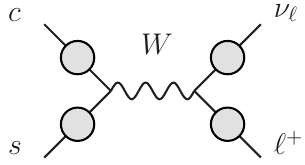}
        \caption{}
    \end{subfigure}  
        \qquad 
    \begin{subfigure}{.3 \textwidth}
        \centering
        \includegraphics[width=1.\linewidth]{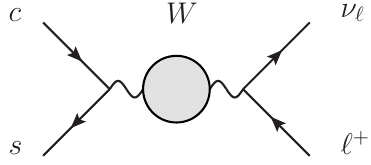}
        \caption{}
    \end{subfigure}    
        \caption{Self-energy diagrams.
        }
        \label{fig:self}
\end{figure}

The $W$ corrections in Fig.~\ref{fig:vertex-1} are equivalent to the ones in the $\mu \to e \nu_\mu \bar \nu_e$ decay. Therefore,  
as long as $G_F$ is determined by the measurement of the muon lifetime, these corrections can be absorbed by a redefinition of $g$. 
The $Z$ corrections are more subtle because the magnitude of the interactions differs between quarks and leptons. 
However, by adding the contributions from self-energy in Fig.~\ref{fig:self}(a), these corrections also become absorbable through the redefinition,
see the next subsection for details.

Next, for the contributions from Fig.~\ref{fig:vertex-2}, 
the summation of them cancels each other out, including divergences. 
In the NDR scheme, the photon corrections in Fig.~\ref{fig:vertex-2} give
\beq
i \mathcal{M}^{\gamma}_{\text{vertex}} & = i \mathcal{M}_0
 \frac{3 \alpha}{4 \pi} \left(Q_c - Q_s + Q_\ell\right)\left[ \frac{1}{\bar{\epsilon}} + \ln\left(  \frac{\mu^2}{M_W^2}\right) + \frac{5}{6} + \mathcal{O}\left(\frac{s}{M_W^2}\right)\right]\non
 &\propto Q_c - Q_s + Q_\ell = 0\,.
\eeq
Similar cancellation occurs for $Z$ corrections:
\beq
\mathcal{M}^Z_{\text{vertex}}
&\propto \left(\frac{1}{2 } -Q_c s_W^2 \right) -
\left(-\frac{1}{2 } -Q_s s_W^2\right)
+  \left(-\frac{1}{2 } -Q_\ell s_W^2 \right)
-\left(\frac{1}{2 }\right)\non
& = 0\,, 
\eeq
where each parenthesis corresponds to the left-handed $Z$ couplings.\footnote{%
For the HV scheme, we have checked that the cancellation of the vertex corrections in Fig.~\ref{fig:vertex-2} is not obvious but still correct. 
In fact, since a contraction of the Dirac matrices:
$\gamma_\mu P_R \gamma^\mu \bar \gamma^\nu P_L = - 2 \epsilon \bar\gamma^\nu P_L $ holds, where $\bar\gamma^\nu$ is the $4$-dimensional gamma matrix
\cite{Belusca-Maito:2020ala}, 
the right-handed $Z$ interactions provide nonzero contributions 
multiplying by ${1}/{\epsilon}$ pole. We find that this contribution is canceled out by the (right-handed) $\gamma$ interaction in the HV scheme. For example, we obtained 
\beq
i \mathcal{M}^{\gamma}_{\text{vertex(b)}} +  i \mathcal{M}^{Z}_{\text{vertex(b)}} & \supset  
 i \mathcal{M}_0
 \frac{3 e}{32 \pi^2} \left[e Q_s +   g_Z \left(- Q_s s_W^2 \right)\frac{c_W}{s_W} \right]\non
 & = 0 \,,
\eeq
from Fig.~\ref{fig:vertex-2}(b). Here, the factor $c_W/s_W$ comes from $WWZ$ vertex.
}

\paragraph{Other contributions.}

Other contributions in Fig.~\ref{fig:self}, except for the virtual-photon self-energy diagram of the charged-lepton leg, do not contribute due to the renormalization, see, \eg, Refs.~\cite{Stuart:1991cc,vanRitbergen:1999fi}.
The self-energy for the charged-lepton leg is classified as the long-distance QED correction and will be calculated in the next section (see Fig.~\ref{fig:LDself}).

\subsection{Corrections to the Fermi constant}
Usually, the Fermi coupling constant
$
G_F= 1.1663788(6) \times 10^{-5} ~\mathrm{GeV}^{-2}
$
is derived from the muon lifetime \cite{ParticleDataGroup:2024cfk}:
\beq
\label{eq:GF}
\tau_\mu^{-1} \equiv \Gamma(\mu \to e \nu_\mu \bar \nu_e (\gamma) )= 
\frac{G_F^2 m_\mu^5}{192 \pi^3} F(\rho)\left(1+ \Delta q \right)\,,
\eeq
with $\rho = m_e^2/m_\mu^2$, through calculations in the Fermi model  where the $W$ propagator is  replaced by the Fermi-contact interaction.
The function $F(\rho)$ represents the electron mass dependence for the phase space integral:
$ F(\rho)=1-8 \rho+8 \rho^3-\rho^4-12 \rho^2 \ln \rho=0.99981295.$
$\Delta q$ represents the QED corrections and has been evaluated up to $\mathcal{O}(\alpha^3)$ terms 
\cite{Kinoshita:1958ru,Nir:1989rm,vanRitbergen:1998yd,vanRitbergen:1999fi,Steinhauser:1999bx,Ferroglia:1999tg,Pak:2008qt,Fael:2020tow,Czakon:2021ybq}.
The leading terms of the QED corrections were derived by Kinoshita and Sirlin \cite{Kinoshita:1958ru}, 
\beq
\tau_\mu^{-1}=\frac{G_F^2 m_\mu^5}{192 \pi^3} F(\rho)\left[1+\frac{\alpha\left(m_\mu\right)}{2 \pi}\left(\frac{25}{4}-\pi^2\right)\right]\,,
\label{eq:mulifetime}
\eeq
which corresponds to the sum of the one-loop virtual photon correction 
and the inclusive contribution from single-real-photon emission with $m_e =0$.
The important point is that the QED corrections $\Delta q$ have been calculated using the optical theorem for the muon self-energy diagrams in the Fermi model,\footnote{It was shown that the IR poles and the singularities as $m_e \to 0$ are canceled in the amplitude level, and the UV divergences are absorbed by the on-shell renormalization in the optical theorem prescription \cite{vanRitbergen:1998yd}.}
so that the resultant QED corrections correspond to the inclusive QED radiative correction, with both hard and soft photon emissions contained.
This fact implies that
all short-distance corrections to the muon lifetime in the full theory have to be conventionally absorbed in the definition of $G_F$.

Therefore, in order to calculate the full one-loop radiative corrections to the meson decays consistently, 
one must also know the complete radiative corrections to the muon lifetime in the full theory as well.
To be more precise, the short-distance QED and EW corrections to the $\mu  \nu_\mu e \nu_e$ operator must be calculated, and we will subtract them from the above results. 
Fortunately, we can readily obtain them
from the calculations in the case of the meson leptonic decays. 

First, $ZW$-box contributions come from the same diagrams as Fig.~\ref{fig:box}, where the strange and charm quark lines are replaced by the muon and muon-neutrino, respectively. 
The result 
can be obtained from  Eq.~\eqref{eq:SDZWbox} by replacing $Q_c \to Q_\nu$ and $Q_s \to Q_\ell$, as
\beq
i\mathcal{M}^{ZW}_{\rm{muon}} & = i\mathcal{M}^{\text{muon}}_0\frac{\alpha(M_Z)}{4\pi} 
  \left[Q_\ell^2 + \frac{5}{2s_W^2}  \left(\frac{1}{s_W^2}   +2 Q_\ell  \right)\right]
\ln{\frac{M_Z^2}{M_W^2}}
\,,
\eeq
with
$\mathcal{M}^{\text{muon}} = - 4 G_F/\sqrt{2}   (\bar{\mu} \gamma^{\mu} P_L \nu_\mu ) (\bar\nu_e \gamma_\mu P_L e ).$
This result is consistent with Refs.~\cite{Marciano:1980pb,Sirlin:1981ie}.

Next, $\gamma W$-box contribution comes from only Fig.~\ref{fig:box}(a) type.
Since the virtual photon correction in $\Delta q$
for the Fermi model corresponds to Fig.~\ref{fig:EFT-one-loop}(a) type, which is already included in Eq.~\eqref{eq:GF}, we have to subtract this contribution. 
So, the short-distance $\gamma W$-box contribution to the muon lifetime is 
\beq
i\mathcal{M}^{\gamma W}_{\rm{muon}}& = 
i \mathcal{M}_{\text{muon(a)}}^{\gamma W} -  \left. i \mathcal{M}_{\text{muon(a)}}^{\gamma }\right|_{\bar{\epsilon}^0}\non 
&=  i \mathcal{M}_0^{\text{muon}} \frac{\alpha}{ 4\pi}  Q_\ell^2 \left[  \ln \left(\frac{ M_W^2}{\mu^2 }\right)+\frac{1}{2} + \kappa_a  \right]\,.
\eeq

Similar to the previous subsection, the photon vertex corrections do not give a contribution to the muon lifetime.

In practice, the other weak radiative corrections also contribute to the muon lifetime.
These radiative corrections are called the universal renormalization factor 
\cite{Marciano:1980pb,Sirlin:1981ie}. 
An important fact is that 
they are also contained within the four-fermion interactions for the meson (semi-)leptonic decays in exactly the same form.
As the most nontrivial check of this cancellation,
we have explicitly calculated the $Z$-boson radiative corrections to the muon/neutrino wave function renormalization constants (in Fig.~\ref{fig:self}(a) type) 
and the muon-neutrino-$Z$ triangle diagram (in Fig.~\ref{fig:vertex-1}(a) type), using the NDR scheme for the UV divergences.
By adding these contributions to the muon decays together, we obtain
\beq
i \mathcal{M}
= - i \mathcal{M}_0^{\text{muon}}\frac{g_z^2}{32\pi^2} \left[1 + \left(Q_\ell - Q_\nu\right) s_W^2 \right]^2 \left[ \frac{1}{\bar{\epsilon}} + \ln\left( \frac{\mu^2}{M_Z^2}\right) - \frac{1}{2}\right]\,,
\eeq
where we discarded the contributions from the $e \nu_e$ side (that is trivially universal compared to the meson (semi-)leptonic decay amplitudes).

On the other hand, in the $D_s^+$ leptonic decays, corresponding virtual-$Z$ contributions can be obtained by replacing $Q_\nu \to Q_c$ and $Q_\ell \to Q_s$.
They lead to the exact same corrections.
Therefore, these weak corrections, including the UV divergence, are classified as the universal renormalization factor, and do not contribute to the $D_s^+$ meson leptonic decay rate \cite{vanRitbergen:1999fi}.

Hence,  
the universal contributions necessarily cancel in the meson (semi-)leptonic decays in total, as long as the Fermi coupling constant is defined by Eq.~\eqref{eq:GF}.
This feature is expected to be independent of the regularization scheme.

\subsection{Summary of the short-distance corrections}

Collecting the contributions obtained above, 
we arrive at the short-distance EW-QED corrections to the $cs\ell\nu$ amplitude:
\beq
\mathcal{M}^{\text{SD}} & =  \mathcal{M}_0+ \mathcal{M}^{ZW} 
+ \mathcal{M}^{\gamma W}
- \mathcal{M}^{ZW}_{\text{muon}}
- \mathcal{M}^{\gamma W}_{\text{muon}}\non
& =  \mathcal{M}_0 \left( 1 + \frac{\alpha(\mu)}{ 4\pi} 
\left\{   Q_\ell \left( - 4 Q_c + Q_s - Q_\ell  \right) \ln \left(\frac{ M_Z^2}{\mu^2 }\right)
+  Q_\ell \left( 5 Q_c + \frac{1}{2}Q_s - \frac{1}{2}Q_\ell \right) \right. \right.\non
& \qquad \qquad 
+\left. Q_\ell \Bigl[ \left(Q_s- Q_\ell\right)  \kappa_a - 4 Q_c \kappa_b\Bigr]\Biggr\}\right)\,.
\label{eq:SDexact}
\eeq
Here, the renormalization group of $\alpha (\mu)$ was not explicitly addressed in our discussion.
Since its scale dependence is small (in view of the precision of the charm physics), we naively take $\alpha (\mu)$ as the overall factor of the short-distance corrections, and two logarithmic terms are merged into a single term \cite{Bigi:2023cbv}.

As mentioned at the beginning of this section, the QCD corrections appear at two-loop level \cite{Sirlin:1977sv,Sirlin:1981ie,Marciano:1983ss,Marciano:1985pd}.
Combining with the QCD correction $(\mathcal{A}_{g_s})$ and  substituting the QED charge in Eq.~\eqref{eq:SDexact}, 
we eventually obtain the short-distance corrections 
\beq
\eta_{\text{EW}} \equiv \frac{\mathcal{M}}{\mathcal{M}_0}&=
\begin{cases}
\displaystyle
1 + \frac{\alpha (\mu) }{ 2\pi} 
\left[  \ln \left(\frac{ M_Z^2}{\mu^2}\right)
+\frac{1}{2}\mathcal{A}_{g_s} 
- \frac{11}{6} \right] & \text{(NDR)}\,,\\[10pt]
 \displaystyle
 1 + \frac{\alpha (\mu)}{2 \pi} 
\left[   \ln \left(\frac{ M_Z^2}{\mu^2}\right)
+\frac{1}{2}\mathcal{A}_{g_s}- \frac{7}{6} \right] & \text{(HV)}\,,
\label{eq:SD-final}
\end{cases}
\eeq
with $\mu = m_{D_s}$ for the $D_s^+$ leptonic decay.\footnote{%
The numerical values are 
\beq
\eta_{\text{EW}} = \begin{cases}
1.0068& \text{(NDR)}\,,\\
1.0076& \text{(HV)}\,,
\label{eq:SD-num}
\end{cases}
\eeq
for the $D_s^+$ leptonic decays.}
In the numerical study in Sec.~\ref{sec:calc}, we use
$\alpha(m_{D^s})^{-1} \simeq \alpha(m_\tau)^{-1} = 133.450(6)$ \cite{ParticleDataGroup:2024cfk}.
For the QCD correction, 
we use a numerical value 
for $\mathcal{O}(1)$\,GeV meson and $\beta$ decays,   
evaluated from Refs.~\cite{Sirlin:1981ie,Marciano:1985pd},\footnote{%
In the pion decays, the QCD correction was estimated as $\mathcal{A}_{g_s} = - (\alpha_s/\pi) \ln (M_Z/m_\rho)$ \cite{Marciano:1993sh}.}
\beq
\mathcal{A}_{g_s} = -0.34\,.
\eeq

The leading-logarithmic term is consistent with the well-known short-distance electroweak correction 
 known as the Sirlin factor \cite{Sirlin:1981ie,Atwood:1989em}.
The $-11/6$ term has been derived in Refs.~\cite{Brod:2008ss,Gorbahn:2022rgl}  and Ref.~\cite{Bigi:2023cbv} independently. 
Both works regularize the photon-induced IR divergences by using dimensional regularization with the NDR scheme and setting all external momenta to zero.
To the best of our knowledge,
this is the first explicit determination of these rational terms using different regularization schemes.
We observed that including these rational terms reduces the leading-logarithmic correction by $24\%$ and $15\%$ in the NDR and HV schemes, respectively. 
It is also noted that the difference between the two regularization schemes is small. 

In standard treatments of QCD corrections,
such scheme dependence at $\mu = M_W$ is canceled by incorporating the two-loop QCD anomalous dimensions of the operators,
while the scheme dependence at the low energy matching scale (the hadronic scale) remains, but can be compensated by the change in $\mu$ \cite{Buchalla:1996fp}.
In our approximate approach, we use a fixed value of $\alpha$ and therefore do not take the QED anomalous dimensions into account.  
If we consider that the difference between two short-distance corrections in Eq.~\eqref{eq:SD-final} can be traded for a change of $\mu$, then choosing 
\beq
\mu_{\text{HV}} =  e^{1/3} \mu_{\text{NDR}} \simeq 1.4 \,\mu_{\text{NDR}} \,
 \eeq
yields identical short-distance corrections in the two schemes.

Furthermore, it is found that the QCD correction $\mathcal{A}_{g_s}$ increases $|V_{cs}|$ by just $2 \times 10^{-4}$.
Since this is an order of magnitude smaller than the numerical value currently under consideration, 
it can be safely neglected.

We should note that the obtained short-distance corrections in Eq.~\eqref{eq:SD-final} 
can be applied to \emph{any} meson leptonic and semi-leptonic decays within the SM.

%=======================================================
%        LONG DISTANCE
%=======================================================
\section{Long-distance QED corrections}
\label{sec:long}
In this section, we compute the long-distance QED corrections to the decays $D_s^+ \to \ell^+ \nu_\ell$
using the point-like approximation, \ie, scalar QED (SQED), for the $D_s^+$ meson.
Based on the definition of the decay constant,
\beq
        \bra{0} \overline{s} \gamma^\mu \gamma_5 c \ket{D_s^+(p)} = i f_{D_s}p^\mu  \,,
\eeq
we matched the effective Lagrangian as follows:
\begin{equation}
    \begin{aligned}
        \mathcal{L}_{\rm eff} =   \mathcal{L}_{\text{SQED}} 
        - \frac{G_F}{\sqrt{2}} V_{cs}^* f_{D_s} \overline{\nu_\ell} \gamma^\mu(1-\gamma_5) \ell D_\mu \phi^+ 
        + \text{h.c.}\,,
    \end{aligned}
\end{equation}
where $\phi^+$ is a complex scalar field that indicates $D_s^+$.
$\mathcal{L}_{\text{SQED}}$ is the corresponding SQED Lagrangian for $\phi^+$, and 
the covariant derivatives are defined as 
\begin{align}
    D_\mu \phi^+      &= (\partial_\mu + i Q_{D_s} e A_\mu) \phi^+\,, 
\end{align}
with $Q_{D_s} = 1.$
Note that this matching produces a dimension-five contact interaction $  \overline{\nu_\ell}\Slash{A} P_L \ell \phi^+$, which is required to obtain the gauge-invariant radiative corrections \cite{Ruderman1958,Decker:1994ea,Kinoshita:2001pn}.  
A global average of the decay constant for $N_f = 2 + 1 + 1$ is $f_{D_s} = 249.9(0.5)$\,MeV  \cite{FlavourLatticeAveragingGroupFLAG:2024oxs}.
Note that the isospin-breaking correction is currently missing in the lattice calculations.

Performing 
direct lattice calculations of the QED corrections to the decay constant
 $f_{D_s}$  remains challenging  \cite{Desiderio:2020oej,Giusti:2023pot,Frezzotti:2023ygt}.
 These works investigated 
the contribution from a single real-photon emission with $E_\gamma > 10$\,MeV, by evaluating hadronic matrix elements with one electromagnetic current insertion.
These QED corrections are classified as the \emph{structure-dependent} part of the real radiations.
It was found that this structure-dependent part is sizable only in the electron mode $D_s^+ \to e^+ \nu \gamma$, 
whereas for $\mu$ and $\tau$ modes,
the QED corrections are dominated by the point-like contribution,
which should correspond to our estimation in the next subsection.
Regarding the structure-dependent part of the virtual corrections, we will comment on them in Sec.~\ref{sec:conc}.

\subsection{Soft photon emissions}
\label{sec:softemission}
\begin{figure}[t]
    \centering
    \begin{subfigure}{.3 \textwidth}
        \centering
        \includegraphics[width=1.\linewidth]{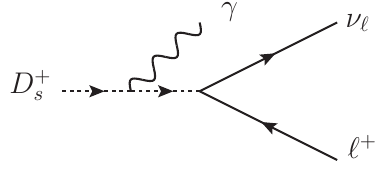}
        \caption{}
    \end{subfigure}
    \quad 
    \begin{subfigure}{.3 \textwidth}
        \centering
        \includegraphics[width=1.\linewidth]{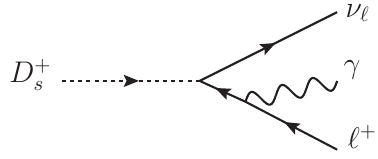}
        \caption{}
    \end{subfigure} 
    \quad 
    \begin{subfigure}{.3 \textwidth}
        \centering
        \includegraphics[width=1.\linewidth]{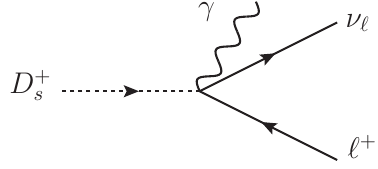}
        \caption{}
    \end{subfigure}      
        \caption{Feynman diagrams for the real photon emissions.
        }
        \label{fig:emission}
\end{figure}

First, we compute corrections from real soft-photon emissions.
As briefly stated in the Introduction, 
for the $\mu$ mode, there is 
a cut that excludes hard photons from the signal 
\cite{BESIII:2021anh,BESIII:2023cym,BESIII:2024dvk}. 
Therefore, we have to analyze the \emph{exclusive} $D_s^+ \to \mu^+ \nu_\mu \gamma$ decay, 
which depends on the energy of the undetected photon.
(We will perform resummation for the missing photons.)
On the other hand, 
for the $\tau$ mode, the situation is different.
Both because the mass difference is small between $D_s^+ $ and $\tau$ lepton,
\beq
m_{D_s} - m_{\tau} \simeq 191\,\text{MeV}\,,
\eeq
and because the decay of $\tau^+$ always produces at least one additional neutrino ($\bar\nu_\tau$), 
the measurements are, in practice, \emph{inclusive} with respect to the soft-photon emissions
\cite{BESIII:2021bdp,BESIII:2021wwd,BESIII:2023ukh,BESIII:2023fhe,BESIII:2024dvk}.
We will discuss these in detail in Sec.~\ref{sec:Emax}.
It is well known that for inclusive observables, the total QED correction is suppressed by the cancellations and its magnitude is expected to be $\mathcal{O}(\alpha/\pi)$, \eg, Eq.~\eqref{eq:mulifetime} (Kinoshita-Lee-Nauenberg theorem \cite{Kinoshita:1962ur,Lee:1964is}).
In contrast, the QED correction can be enhanced in the exclusive decays by the large logarithmic terms that arise from significant differences among  
the small undetected photon energy, the light lepton mass, and the meson masses. 
So, we should expect that the total QED correction is significant in the $\mu$ mode, but not in the $\tau$ mode.

Diagrams of the (single) photon emissions (with the momentum $k$), which are also known as inner-bremsstrahlung,   are shown in Fig.~\ref{fig:emission}, 
and these matrix elements can be simplified as follows:  
\beq
i \mathcal{M}_{\text {real }}^{\text{(a)}} & =- i\mathcal{M}_0 e Q_{D_s} \frac{p_{D_s} \cdot \epsilon^*(k)}{p_{D_s} \cdot k - i \epsilon } \,,\\
%%%%%
i \mathcal{M}_{\text {real }}^{\text{(b)}} 
& =-i\mathcal{M}_0  e Q_\ell \frac{ p_\ell \cdot \epsilon^*(k)}{p_\ell \cdot k +  i \epsilon }  \non
& \quad
- e Q_\ell \frac{G_F}{\sqrt{2}} V_{cs}^* f_{D_s}  \overline{u}_{\nu_\ell}(1+\gamma_5) \left( \frac{m_\ell \slashed{k}}{2p_\ell \cdot k +  i \epsilon} - 1  \right) \slashed{\epsilon}^* (k)v_\ell \,,\label{eq:fsr}\\
%%%%%
i \mathcal{M}_{\text {real }}^{\text{(c)}} & =e Q_{D_s}\frac{G_F}{\sqrt{2}} V_{c s}^* f_{D_s} \overline{u}_{\nu_\ell}\left(1+  \gamma_5\right)\Slash{\epsilon}^* (k) v_\ell \,,
\eeq
where the tree-level amplitude $\mathcal{M}_0$ is defined by
\beq
\mathcal{M}_0 = -i \frac{G_F}{\sqrt{2}} 
V_{c s}^* f_{D_s} m_\ell \overline{u}_{\nu_\ell}  \left( 1+\gamma_5 \right) v_\ell\,.\label{eq:tree}
\eeq
The sum of these three amplitudes is \cite{Bryman:1982et}
\beq
i \mathcal{M}_{\text {real }}
& = -i \mathcal{M}_0  e\left( Q_{D_s}\frac{p_{D_s} \cdot \epsilon^*}{p_{D_s} \cdot k - i \epsilon }+ Q_\ell \frac{p_\ell \cdot \epsilon^*(k)}{p_\ell \cdot k + i \epsilon}\right)
\non
& \quad
- e Q_\ell \frac{G_F}{\sqrt{2}} V_{c s}^* f_{D_s}  \frac{m_\ell }{2 p_\ell \cdot k + i \epsilon} \overline{u}_{\nu_\ell}\left(1+\gamma_5\right) \Slash k \Slash{\epsilon}^\ast v_\ell\,.
\label{eq:IBtotal}
\eeq
Although the matrix element $\mathcal{M}_{\text {real }}^{\text{(c)}} $ is canceled by part of $\mathcal{M}_{\text {real }}^{\text{(b)}}$ in the real photon emission amplitude, this vertex must be kept for the full virtual correction; see the next subsection.
The first line in Eq.~\eqref{eq:IBtotal} collects the leading soft-photon contribution containing the infrared (IR) divergence, whereas the second line is a subleading term whose contribution is significant only for hard photons.
Because the total amplitude of the inner-bremsstrahlung is still proportional to the lepton mass, it receives the helicity suppression.

Taking the spin and polarization sums for the final state particles, we obtain
\begin{align}
        \sum_{s, \lambda}
        \left|\mathcal{M}_{\text {real}}\right|^2 
        &=
        4 G_F^2\left|V_{c s}\right|^2 f_{D_s}^2 m_\ell ^2 e^2  \non
        & \hspace{0.5em} \times \Biggl\{ - \left(p_\ell \cdot p_{\nu_\ell}\right)\left[Q_{D_s}Q_\ell\frac{2 p_{D_s} \cdot p_\ell}{\left(p_{D_s} \cdot k\right)\left(p_\ell \cdot k\right)}+ Q_{D_s}^2\frac{m_{D_s}^2}{\left(p_{D_s} \cdot k\right)^2}
        + Q_\ell^2 \frac{m^2_\ell}{\left(p_\ell \cdot k\right)^2}\right] \non
        &\qquad  +Q_{\ell}^2\frac{p_{\nu_\ell} \cdot k}{p_\ell \cdot k}-Q_{\ell}^2\frac{m^2_\ell \left(p_{\nu_\ell} \cdot k\right)}{\left(p_\ell \cdot k\right)^2}-Q_{\ell}Q_{D_s}\frac{p_\ell \cdot p_{\nu_\ell}}{p_\ell \cdot k} \non
        &  \qquad  +Q_{\ell}Q_{D_s}\frac{p_{D_s} \cdot p_{\nu_\ell}}{p_{D_s} \cdot k}-Q_{\ell}Q_{D_s}\frac{\left(p_{D_s} \cdot p_\ell\right)\left(p_{\nu_\ell} \cdot k\right)}{\left(p_{D_s} \cdot k\right)\left(p_\ell \cdot k\right)}\Biggr\} \,, 
        \label{eq:photon_emission}
\end{align}
where the $\pm i \epsilon$ terms have been omitted for clarity. 
The terms in the square bracket 
contain the IR singularity.
For the analytic phase space integral,
we use the Dalitz plot formalism, see Appendix~\ref{sec:Dalitz}.
 In the IR limit, which corresponds to the  two-body decay phase space integral for $\ell^+ \nu_\ell$, 
 it is hard to perform the Dalitz integration, so we use the soft-photon approximation  \cite{Yennie:1961ad, Weinberg:1965nx,Isidori:2007zt}, namely:
\beq
\int_{E_\gamma < E_{\rm max}}
\frac{d^3 \bm{k}}{(2\pi)^3}\frac{1}{2 E_\gamma} \int d \Pi_2 = 
\frac{1}{8\pi^2}
\int_{m_\gamma}^{E_{\rm max}} d E_\gamma
\sqrt{E_\gamma^2  - m_\gamma^2}\int^1_{-1} d \cos \theta \int d \Pi_2 \,,
\label{eq:softapp}
\eeq
where $m_\gamma$ is the soft photon mass introduced to regulate the IR divergences. $\int d \Pi_2$ is the usual two-body phase space.
Here, $E_{\rm max}$
 represents the maximum energy of the undetected soft photon in the $D_s^+$ rest frame.
After performing the three-body analytic integrations \cite{ParticleDataGroup:2024cfk},
we obtain the following QED corrections from the single photon radiation:
\begin{align}
    \left.  \frac{\Gamma(D_s^+\to \ell^+\nu_\ell \gamma)}{\Gamma_0}\right|_{E_\gamma < E_{\rm max}}
        & =\frac{\alpha_0}{2\pi}\Biggl\{-4\left[1+\frac{1+x_\ell}{2(1-x_\ell)}\ln{x_\ell}\right]\ln{\left(\frac{2E_{\rm{max}}}{m_\gamma}\right)}
        \non
        &\quad 
        +2  - \frac{1+x_\ell}{1-x_\ell}\left[\ln{x_\ell}+ \frac{1}{2}\ln^2{x_\ell}+ 2 \text{Li}_2(1 - x_\ell) \right]
        \non 
        & \quad +F^\text{hard}\left(E_\text{max}\right) \Biggr\}
        \,, \label{eq:result_photon_emission}
\end{align}
with the tree-level partial decay width 
\beq
\Gamma_0 =\frac{G_F^2}{8 \pi}\left|V_{c s}\right|^2  m_{D_s} m_\ell^2 f_{D_s}^2\left(1-x_\ell\right)^2
\,, 
\eeq
and the hard-photon correction 
\beq
F^\text{hard}\left(E_\text{max}\right)&=
\frac{\alpha_0}{2\pi}\Bigg\{\frac{E_{\rm{max}}}{m_{D_s}(1-x_\ell)^2}\left[11-14x_\ell-\frac{3E_{\rm{max}}}{m_{D_s}}+4(1-x_\ell)\ln{x_\ell}-\frac{2E_{\rm{max}}}{m_{D_s}}\ln{x_\ell}\right]\notag\\
&\qquad +\frac{1}{(1-x_\ell)^2}\left[\frac{3}{2}-3x_\ell-4(1-x_\ell)\frac{E_{\rm{max}}}{m_{D_s}}+\frac{2E_{\rm{max}}^2}{m_{D_s}^2}\right]\ln{\left(1-\frac{2E_{\rm{max}}}{m_{D_s}}\right)}\non
& \qquad -2\left(\frac{1+x_\ell}{1-x_\ell}\right)\text{Li}_2\left(\frac{2E_{\rm{max}}}{m_{D_s}}\right)\Bigg\}\,,
\label{eq:Fhard}
\eeq
where $x_\ell=m_\ell^2/m_{D_s}^2$.
Here, for the long-distance QED corrections,
we use the fine-structure constant $\alpha_0^{-1} = 137.035999178(8)$ \cite{ParticleDataGroup:2024cfk}.
The first line of Eq.~\eqref{eq:result_photon_emission} gives $\mathcal{O}(\ln E_{\rm max})$ contributions (including the IR singularity), and the second line provides a constant correction as $\mathcal{O}((E_{\rm max})^0)$. 
The terms in the square bracket in Eq.~\eqref{eq:photon_emission} with 
the soft-photon approximation in Eq.~\eqref{eq:softapp} precisely provide these two contributions.
On the other hand, 
the third line of Eq.~\eqref{eq:result_photon_emission}, $F^{\text{hard}}$, corresponds to $\mathcal{O}(E_{\rm max})$ contribution, which is usually discarded in the soft-photon approximation. 
Note that 
$F^{\text{hard}}$ vanishes in the soft photon limit, $F^{\rm hard}(E_{\rm max} \to 0) = 0$.
We will show later that the contributions from $F^{\text{hard}}$ are tiny. 
(In Sec.~\ref{sec:Emax}, we will provide a detailed discussion of
the dependence of $E_{\text{max}}$ and 
how $E_{\text{max}}$ is handled in the experimental analyses.)
The analytic formula of $F^{\text{hard}}$ is consistent with Eq.~(50) of Ref.~\cite{Carrasco:2015xwa}.\footnote{We appreciate Toru Goto for an independent check of the formulae in this subsection.}

As a nontrivial check, we have confirmed the inclusive limit of photon emission, $E_{\text{max}} = (m_{D_s}^2 - m_\ell^2)/2 m_{D_s} $. We found that our result is consistent with a famous result by Kinoshita for the fully inclusive inner-bremsstrahlung process of $\pi^+ \to e^+ \nu_e \gamma$~\cite{Kinoshita:1959ha} 
(for details, see Appendix~\ref{sec:inclusive}).

\subsection{Virtual corrections}

\begin{figure}[t]
    \centering
    \begin{subfigure}{.3 \textwidth}
        \centering
        \includegraphics[width=1.\linewidth]{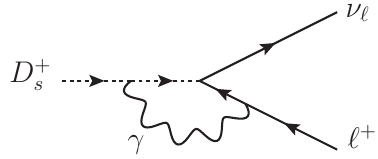}
        \caption{}
    \end{subfigure}
    \quad 
    \begin{subfigure}{.3 \textwidth}
        \centering
        \includegraphics[width=1.\linewidth]{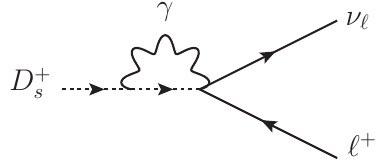}
        \caption{}
    \end{subfigure} 
    \quad 
    \begin{subfigure}{.3 \textwidth}
        \centering
        \includegraphics[width=1.\linewidth]{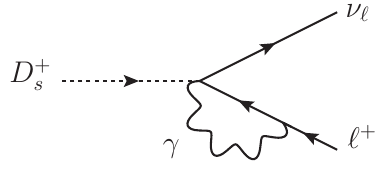}
        \caption{}
    \end{subfigure}      
        \caption{Virtual vertex corrections for the long-distance QED correction.
        }
        \label{fig:virtual}
\vspace{0.6cm}
    \begin{subfigure}[t]{.3 \textwidth}
        \centering
        \includegraphics[width=1.\linewidth]{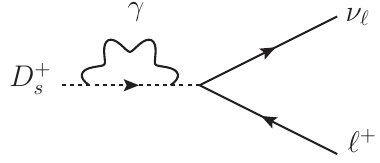}
        \caption{}
    \end{subfigure}  
        \qquad 
    \begin{subfigure}[t]{.3 \textwidth}
        \centering
        \includegraphics[width=1.\linewidth]{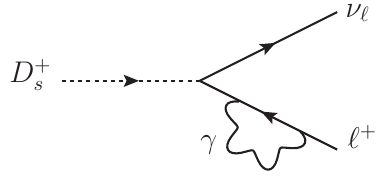}
        \caption{}
    \end{subfigure}    
        \caption{Self-energy diagrams.
        }
        \label{fig:LDself}
\end{figure}

Next, we calculate the virtual corrections to $D_s^+\to\ell^+\nu_\ell$, which are comprised of the vertex corrections in Fig.~\ref{fig:virtual} and the self-energy diagrams in Fig.~\ref{fig:LDself}.
Both contributions contain the IR singularities, and they must be canceled out when added to the contributions from the soft-photon emissions, as explicitly shown in the next subsection.
Regulating the singularities by the soft-photon mass,  
we obtained each result for the vertex corrections, 
%%%
\begin{align}
\frac{\Delta \Gamma_{(\text{vertex})}^{\text{(a)}}}{\Gamma_{\text{0}} } 
  &=- Q_\ell Q_{D_s}\frac{\alpha_0}{2\pi}\Biggl[\frac{5}{2}\left( \frac{1}{\bar{\epsilon}} + \ln  \frac{\mu^2}{m_{D_s}^2}\right)+\frac{11}{2}   + \kappa_c \non
  & \qquad \qquad  \quad
  -2\left(\frac{1+x_\ell}{1-x_\ell}\right)\ln{x_\ell}\ln{\frac{m_\gamma}{\sqrt{m_{D_s}m_\ell}}} -\frac{2}{1-x_\ell}\ln{x_\ell}\Biggr]\,,\\
  \frac{\Delta\Gamma^{\text{(b)}}_{(\text{vertex})}}{
  \Gamma_{\text{0}}} &=    Q_{D_s}^2 \frac{\alpha_0}{2\pi} \left[ -\frac{3}{2} \left( \frac{1}{\bar{\epsilon}} + \ln \frac{\mu^2}{m_{D_s^2}} \right) -\frac{7}{2} \right] \,,\\
  \frac{\Delta\Gamma^{\text{(c)}}_{(\text{vertex})}}{\Gamma_{\text{0}} }&=  - Q_\ell Q_{D_s}\frac{\alpha_0}{2\pi} \left[ -3\left( \frac{1}{\bar{\epsilon}} + \ln \frac{\mu^2}{m_\ell^2} \right) -4  + \kappa_d \right]\,,
\end{align}
and for the wave-function renormalization constants, 
\begin{align}
  \frac{\Delta\Gamma^{\text{(a)}}_{(\text{self})}}{
  \Gamma_{\text{0}}} &=    2 \left( \frac{1}{2}\left. \frac{d \Sigma}{dp^2}\right|_{p^2 = m_{D_s}^2} \right) \non
      &=Q_{D_s}^2
      \frac{\alpha_0}{2\pi}\left(\frac{1}{\bar{\epsilon}}+\ln\frac{\mu^2}{m_{D_s}^2}-2\ln\frac{m_\gamma}{\sqrt{m_{D_s}m_\ell}}-\frac{1}{2}\ln{x_\ell}\right)\,,\\
%%%%
    \frac{\Delta\Gamma^{\text{(b)}}_{(\text{self})}}{
  \Gamma_{\text{0}}} &=     2 \left(
  \frac{1}{2}\left.\frac{d\Sigma}{d\slashed{p}}\right|_{\slashed{p} = m_\ell}\right)\non  
      &=Q_{\ell}^2\frac{\alpha_0}{2\pi}\left[-\frac{1}{2} \left(\frac{1}{\bar{\epsilon}} + \ln\frac{\mu^2}{m_\ell^2} \right)-2-2\ln\frac{m_\gamma}{\sqrt{m_{D_s}m_\ell}}+ \frac{1}{2}\ln{x_\ell}\right]\,,
\end{align}
where $\Sigma$ denotes the self-energy amplitudes 
for the external legs. 
To regulate UV divergence, we use dimensional regularization, with $\kappa_c$ and $\kappa_d$ as regularization-scheme-dependent parameters. 
Again, in the same manner as short-distance calculations, we parametrized $\kappa_c = \kappa_d = 0$ in the NDR scheme,\footnote{%
The result of the virtual corrections is consistent with, \eg, Ref.~\cite{Rowe:2024pfs}} while $\kappa_c =1 $ and $\kappa_d = -1$ in the HV scheme.
Note that $\Gamma^{\text{(b)}}_{(\text{vertex})}$ and the wave-function renormalization constants are common to both schemes. 
Here, we use the on-shell renormalization, and the renormalized masses are set equal to their physical masses. 

The renormalization scale $\mu$ 
 should be chosen at the matching scale to the UV theory, 
 namely, the characteristic scale of the structure-dependent radiative corrections in the present study.
 Because such effects are not incorporated in this paper, the remaining $\mu$-dependence is treated as theoretical uncertainty for the extraction of  $|V_{cs}|$, and we drop the $1/\bar{\epsilon}$ terms.

Unexpectedly, we find that the contributions from $\kappa_c$ and $\kappa_d$ are canceled in the HV scheme. 
Hence, total virtual corrections are scheme-independent between the NDR and HV schemes.

\subsection{Summary of the long-distance corrections}

Combining the contributions from the single photon radiations and the virtual corrections, we derive the long-distance QED corrections to $\Gamma(D_s^+ \to \ell^+ \nu_\ell)$, and classify them into the corrections from the initial state radiation (ISR), final state radiation (FSR), interference contributions between the ISR and FSR (INT), hard-photon contributions (hard), and other virtual corrections, as follows: 
\beq
\frac{\Gamma(D_s^+ \to \ell^+ \nu_\ell)^{\rm{LD}}}{\Gamma_0}
&= 1 
+ \frac{\Delta \Gamma_{(\text{vertex})}^{\text{(a}+\text{b}+\text{c)}}}{\Gamma_{\text{0}} } +  \frac{\Delta\Gamma^{\text{(a}+\text{b)}}_{(\text{self})}}{
  \Gamma_{\text{0}}} +  \left.  \frac{\Gamma(D_s^+\to \ell^+\nu_\ell \gamma)}{\Gamma_0}\right|_{E_\gamma < E_{\rm max}}\non
& = 
1+ F^{\text{ISR}}+F^{\text{FSR}}+F^{\text{INT}}+F^{\text{hard}} + H^{\text{(a)}}+H^{\text{(b}+\text{c)}}\,,
\label{eq:longdef}
\eeq
where corrections $F$ represent the real photon emissions, while $H$ represent the virtual corrections. (The corrections from the wave-function renormalization constants are included in $H^{\text{(a)}}$.) 
Each contribution is infrared-safe and is defined in a finite manner\footnote{Both the real photon emissions and the virtual corrections initially have the IR singularities.
By transposing the terms proportional to $\ln (m_\gamma/\sqrt{m_{D_s}m_\ell})$ from the virtual corrections to the real photon emissions, we cancel the singularities and express finite results.} as follows:
\beq
F^{\text{ISR}}&=\frac{\alpha_0}{2\pi}\left( -2\ln{\frac{2E_{\rm{max}}}{\sqrt{m_{D_s}m_\ell}}}+2\right)\,,\\
F^{\text{FSR}}&=\frac{\alpha_0}{2\pi}\left(-2\ln{\frac{2E_{\rm{max}}}{\sqrt{m_{D_s}m_\ell}}}-\frac{1+x_\ell}{1-x_\ell}\ln{x_\ell}\right)\,,\\
F^{\text{INT}}&=\frac{\alpha_0}{2\pi}\left\{ 
-2\left(\frac{1+x_\ell}{1-x_\ell}\right)\ln{x_\ell}\ln{\frac{2E_{\rm{max}}}{\sqrt{m_{D_s}m_\ell}}}-\frac{1+x_\ell}{1-x_\ell}\left[ \frac{1}{2}\ln^2{x_\ell}+2\text{Li}_2(1-x_\ell)\right] \right\}\,,\\
H^{\text{(a)}}
&=\frac{\alpha_0}{2\pi}\left(
3 \ln \frac{\mu^2}{m_{D_s} m_\ell}
-\frac{2 x_\ell }{1-x_\ell}\ln{x_\ell}+\frac{7}{2} \right)\,,\\
H^{\text{(b}+\text{c)}}
&=\frac{\alpha_0}{2\pi}\left(-\frac{9}{2}\ln{\frac{\mu^2}{m_{D_s} m_\ell}} + \frac{3}{4} \ln x_\ell -\frac{15}{2}\right)\,,
\eeq
and $F^{\text{hard}}$ is defined in Eq.~\eqref{eq:Fhard}.
Note that these results are the same in both the NDR and HV schemes, as discussed in the previous subsection.

Furthermore, we carry out a resummation of the potentially large contributions of $\mathcal{O}(\alpha^n \ln^n E_{\rm max}) $, which come from the emission of $n$ soft photons.
This resummation is also required to remove the $E_{\rm max} \to 0 $ singularity.
The resummation of the soft-photon emissions 
is accomplished through the replacement of the following terms with an exponential factor $\Omega_B$
\cite{Weinberg:1965nx,Isidori:2007zt,deBoer:2018ipi},
\beq
1 + F^{\text{ISR}} + F^{\text{FSR}}+F^{\text{INT}} & \ni 1+\frac{\alpha_0}{2\pi}\left[ -4
-2 \left(\frac{1+x_\ell}{1-x_\ell}\right) \ln{x_\ell}\right]\ln{\left(\frac{2E_{\rm{max}}}{\sqrt{m_{D_s}m_\ell}}\right)}\non
&\xrightarrow{\text{resum.}} \left(\frac{2E_{\rm{max}}}{\sqrt{m_{D_s}m_\ell}}\right)^{-\frac{2\alpha_0}{\pi}\left( 1+\frac{1}{2} \frac{1+x_\ell}{1-x_\ell}\ln x_\ell\right)}\non
& \equiv \Omega_B(E_{\text{max}})\,.
\eeq
Here, $E_{\rm max}$ should be reinterpreted as the maximum total energy of undetected soft photons.
 We consider this resummation effect when $|V_{cs}|$ is extracted from the data.
We will present a numerical comparison of the individual contributions of the long-distance QED corrections in the next section.

%=======================================================
%        RESULT
%=======================================================
\section{Complete radiative corrections to \texorpdfstring{\boldmath{$D_s^+ \to \ell^+ \nu$}}{Ds+ to l+ nu}}
\label{sec:calc}

Finally, we obtain the complete one-loop QED corrections to $D_s^+ \to \ell^+ \nu_\ell$ by combining the short- and long-distance corrections in the NDR scheme,
\beq
\Gamma\left(D_s^{+} \rightarrow \ell^{+} \nu_{\ell}\right) 
& =
\Gamma_0 \left\{1 + 
\frac{\alpha(m_{D_s})}{2 \pi}\left[\ln \left(\frac{M_Z^2}{m_{D_s}^2}\right)+\frac{1}{2} \mathcal{A}_{g_s}-\frac{11}{6}\right]
\right\}^2 \non
&\quad  \times \Omega_B
\Biggl\{
1 + 
\frac{\alpha_0}{2 \pi}\left[-2 -\frac{1+x_{\ell}}{1-x_{\ell}}\left(\frac{1}{2} \ln ^2 x_{\ell}+2 \operatorname{Li}_2\left(1-x_{\ell}\right)\right)\right.\non
& \qquad \qquad \left.- \frac{3}{2} 
\ln \frac{\mu^2}{m_{D_s} m_{\ell}}
- \frac{1 + 15 x_\ell}{4 (1 - x_\ell)} \ln x_{\ell}\right]
+
F^{\text{hard}}
\Biggr\}\,.
\label{eq:Gammamaster}
\eeq
In the HV scheme, the term of $-11/6$ in the first line
is replaced by $
-7/6$ (see Eq.~\eqref{eq:SD-final}).
Note that the structure-dependent QED radiative corrections are currently missing from this formula, which will be discussed in Sec.~\ref{sec:conc}.
In this section, we extract the state-of-the-art value of $|V_{cs}|$
from the latest data, and consider the CKM unitarity test.

\subsection{Estimation of \texorpdfstring{$E_{\rm max}$}{Emax}}
\label{sec:Emax}

Before examining the numerical details, 
we first discuss the experimental circumstances for the value of $E_{\text{max}}$, which is a crucial parameter in the numerical calculation of the long-distance QED corrections.
In the experiments,
the threshold energy of the electromagnetic calorimeter to detect the photons is $\mathcal{O}(10)$\,MeV, in order to tag the $\pi^0 \to 2\gamma$ events.
But, this value does not correspond to $E_{\rm max}$.
In practice, since the beamline radiates a large number of photons, 
one has to isolate the signal by applying dedicated cuts that effectively ignore many visible soft photons.
Although different experiments adopt different selection cuts,
here we consider one of the cuts used by the  
BES III Collaboration, since 
they provide the most precise measurements of the $D_s^+$ decays.

To collect the signal of $D_s^+ \to \ell^+ \nu_\ell$, 
it is useful to adopt the missing-squared-mass cut, which is defined as
\beq
M_{\rm miss}^2 = E_{\rm miss}^2 - \left| \vec{0} - \vec{p}_{\rm tag}  - \vec{p}_{\gamma(\pi^0)}-\vec{p}_\mu\right|^2\,,
\eeq
with
\beq
E_{\rm miss} = E_{\rm cm} - E_{\rm tag} - E_{\gamma(\pi^0)} - E_{\mu}\,,
\eeq
where $\gamma(\pi^0)$ represents a monochromatic photon from $D_s^{\ast +} \to D_s^+ \gamma $ and two photons from $D_s^{\ast +} \to D_s^+ \pi^0 \to  D_s^+ \gamma \gamma$.
$M_{\rm miss}^2$ is Lorentz invariant, while the RHS is defined in the laboratory frame.

In the $\ell=\mu $ channel,
if there are no soft-photon radiations,
this reduces to the invariant mass of a single neutrino and therefore must vanish, \ie, $M_{\rm miss}^2= m_{\nu_\mu}^2 = 0$.
The relation between the missing-squared-mass and the single soft-photon momentum in the $D_s^+$ rest frame is \cite{deBoer:2018ipi}
\beq
M_{\text{miss}}^2 &=(p_{\nu_\mu}+p_\gamma)^2\non
& =2 E_{\nu_\mu} E_\gamma(1-\cos{\theta_{\nu\gamma}})\,,
\label{eq:missmu}
\eeq
where $\theta_{\nu\gamma}$ is an undetectable angle between emitted soft-photon and $\nu_\mu$.
In the BES III experiments, 
to avoid a peaking background from $D_s^+ \to K_0 \pi^+$ with missing $K_0$,  
a hard cut $M_{\rm miss}^2 < 0.2\,\text{GeV}^2$ has been set \cite{BESIII:2021anh},\footnote{%
The real photon radiations from $D_s^+ \to \mu^+ \nu_\mu$ do not affect 
another hard cut,
\beq 
\Delta E \in (-0.05, 0.10)\,\text{GeV}\,,
\eeq
where 
$\Delta E = E_{\rm cm} - E_{\rm tag}  - E_{\gamma(\pi^0)} -\sqrt{m_{D_s}^2 + |\vec{0} - \vec{p}_{\rm tag}  - \vec{p}_{\gamma(\pi^0)}|^2} $ \cite{BESIII:2023cym}. 
Here, $\sqrt{m_{D_s}^2 + |\vec{p}_{\rm tag} + \vec{p}_{\gamma(\pi^0)}|^2}$ corresponds to the energy of $D_s^+$ immediately after the production prior to the soft-photon emissions, so that the soft photons from the $D_s^+$ decays do not change $\Delta E$.} 
which leads to
\beq
E_\gamma E_{\nu_\mu} (1-\cos{\theta_{\nu\gamma}}) < 0.1 \text{~GeV}^2\,.
\label{eq:missbound}
\eeq
In the soft-photon limit, the energy of the muon-neutrino is 
\beq
E_{\nu_\mu} = \frac{m_{D_s}}{2} \left(1 - \frac{m_\mu^2}{m_{D_s}^2}\right) \qquad \text{for~} E_\gamma =0\,.
\label{eq:neutrinoE}
\eeq

\begin{figure}[t]
\begin{center}
    \includegraphics[width=0.8\textwidth]{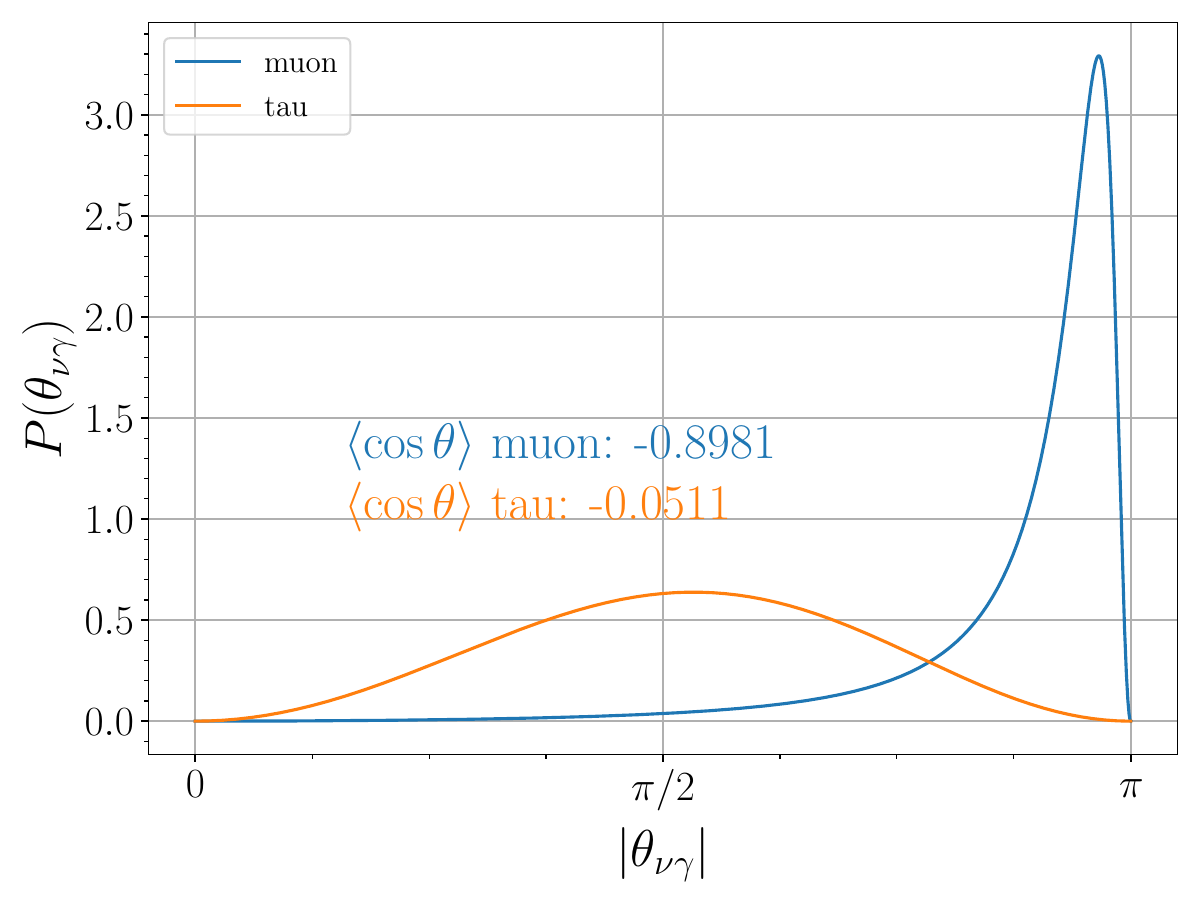}
    \caption{Probability density function of angle between the neutrino and soft photon.} 
    \label{fig:PDF}
\end{center}
\end{figure}

For evaluation of magnitude of $\cos{\theta_{\nu\gamma}}$, 
we consider its expected value. 
Using the soft-photon approximation,
the phase space integration is proportional to (see Eq.~\eqref{eq:photon_emission})
\begin{align}
    I = \int \frac{d^3 \bm{k}}{(2\pi)^3}\frac{1}{2E_\gamma} \left[ \frac{2p_{D_s} \cdot p_\ell}{(p_{D_s}\cdot k)(p_\ell\cdot k) } - \frac{m_{D_s}^2}{(p_{D_s}\cdot k )^2} - \frac{m_\ell^2}{(p_\ell\cdot k)^2}
    \right]\,.
\end{align}
Exploiting the fact that the lepton and neutrino are emitted almost back to back for the soft-photon process, we use $(\pi - \theta_{\nu\gamma})$ for the angle between the lepton and soft-photon.
Then, the integral can be simplified to 
\beq
I \propto \int d\Omega \left[ \frac{2}{1+\beta_\ell\cos\theta_{\nu\gamma}}-1-\frac{1-\beta_\ell^2}{(1+\beta_\ell\cos\theta_{\nu\gamma})^2}\right]\,,
\eeq
with 
\beq
\beta_\ell=\frac{1-x_\ell}{1+x_\ell}\,.
\eeq
This solid-angle integrand gives the (unnormalized) probability density function of the parameter $\theta_{\nu\gamma}$:
\beq
P(\theta_{\nu\gamma})=\frac{2}{1+\beta_\ell\cos\theta_{\nu\gamma}}-1-\frac{1-\beta_\ell^2}{(1+\beta_\ell\cos\theta_{\nu\gamma})^2}\,.
\eeq
In Fig.~\ref{fig:PDF}, 
we show the $P(\theta_{\nu\gamma})$ for the $\mu$ and $\tau$ modes.
%%%%%%%%%%%%%%%%%%%%%% 
%%%%%%%%%%%%%%%%%%%%%% 
From the probability density functions, we obtain the expected values:
\beq
\begin{aligned}
\langle\cos \theta_{\nu\gamma}\rangle &= -0.8981 \qquad \text{for~$\mu$~mode}\,,\\
\langle\cos \theta_{\nu\gamma}\rangle &= -0.0511 \qquad \text{for~$\tau$~mode}\,.
\label{eq:expectedangle}
\end{aligned}
\eeq
It clearly presents that 
in the $\mu$ mode, the soft photon tends to be emitted in the direction of the muon, whereas in the $\tau$ mode, it tends to be emitted in a direction different from that of the tau-neutrino and tau.

Using Eqs.~\eqref{eq:neutrinoE} and \eqref{eq:expectedangle}, 
for the $\mu$ mode we obtain a numerical expected value of $E_{\text{max}}$
as
\beq
E_\gamma < E_\text{max} \simeq 
54\ \text{MeV}& \qquad (\text{for~}\mu \text{~mode})\,.
\label{eq:Emax-num}
\eeq
Note that this $ E_\text{max}$ is calculated in the $D_s^+$ rest frame, which is precisely consistent with what we discussed in
Sec.~\ref{sec:long} for the long-distance QED corrections.

On the other hand, for the $\tau$ mode, the estimation of $E_\text{max}$ is more involved, because the invisible system has to consist of at least tau-neutrino, its anti-neutrino and possibly another neutrino,  and the simple relation in Eq.~\eqref{eq:missmu} no longer holds.
Furthermore, the cuts are entirely different depending on the $\tau^+ $ decay channels \cite{BESIII:2021bdp,BESIII:2021wwd,BESIII:2023ukh,BESIII:2023fhe,BESIII:2024dvk}.

The most precise measurement is given via the $\tau^+ \to e^+ \nu_e \bar{\nu}_{\tau}$ channel for the $\tau$ decays \cite{BESIII:2021bdp}. 
In this measurement, the signals are tagged in an almost inclusive way with respect to soft photons,\footnote{%
Here, events with an extra electromagnetic-calorimeter-shower energy smaller than $0.4$\,GeV are selected. This is a less stringent cut compared to the maximum value of $E_\gamma$: $(m_{D_s}/2) (1 - m_\tau^2/m_{D_s}^2) = 182$\,MeV.} 
excluding within a $5^\circ$ cone of the positron direction. 
The second most precise measurement is given by ${\ensuremath{\tau}}^+\to {\ensuremath{\pi}}^+ \bar{\nu}_\tau$ channel with
 the hard cut $M_{\rm miss}^2 < 0.6\,\text{GeV}^2$ \cite{BESIII:2023fhe}.
In this case,  we expect
\beq
M_{\text{miss}}^2 &=(p_{\nu_\tau} + p_{\nu_{\bar\tau}} +p_\gamma)^2\non
& \simeq 2 E_{\nu_{\bar\tau}} \left(E_\gamma + E_{\nu_\tau}\right)\,.
\eeq
Here, a flat angular distribution is assumed for the soft-photon direction
according to Fig.~\ref{fig:PDF}.
Very naively, $2 E_{\nu_{\bar\tau}} \simeq m_\tau$ holds because ${\ensuremath{\tau}}^+\to {\ensuremath{\pi}}^+ \bar{\nu}_\tau$ is two-body decay. These relations give
\beq
E_\gamma + E_{\nu_\tau} \lesssim   340\,\text{MeV}\,.
\eeq
Since this cut is less stringent than the maximum soft-photon energy $E_\gamma < 182$\,MeV, we conclude that this channel is also collected inclusively.
Note that  $E_{\rm max}$ dependence is very weak in the long-distance QED corrections for the $\tau$ mode, which will be shown in Sec.~\ref{seq:result}.

\subsection{PHOTOS QED corrections}
In many collider experiments, including BESIII,
the PHOTOS Monte-Carlo generator has been used 
for the simulation of modifications only of the kinematic variables  induced by the FSR in the leading-logarithmic approximation \cite{Barberio:1990ms,Barberio:1993qi,Golonka:2005pn,Davidson:2010ew}.
The PHOTOS can be applied in a model-independent way even when the underlying amplitudes are not known,
whereas the theoretical calculations are performed using the specific amplitudes derived from the (effective) models.
This also means that a radiation from the initial leg and interference between the ISR and FSR are not included in the PHOTOS.
Because some experimental results have already been corrected using the PHOTOS,
the theoretical prediction has to be adjusted accordingly when assessing the precise impact of the QED corrections.
Namely, 
to avoid double counting of the QED corrections, one subtracts the FSR contribution from the prediction of the long-distance QED corrections.
In particular,
one has to subtract the FSR contribution from the charged-lepton leg in the $D_s^+ \to \ell^+ \nu_\ell$ process.

The FSR contribution in the leading-logarithmic approximation comes from the soft-photon emission amplitude in Eq.~\eqref{eq:fsr},
\beq
    i\mathcal{M}_{0}  e  \frac{p_\ell\cdot \epsilon^*(k)}{p_\ell \cdot k + i \epsilon} \,,
\eeq
where the tree-level amplitude $\mathcal{M}_0$ is given in Eq.~\eqref{eq:tree}.
%with the self-energy correction
The corresponding virtual correction from the charged-lepton leg comes from the leading-logarithmic approximation of the self-energy correction,
\begin{align}
   \left. \frac{d\Sigma}{d\slashed{p}}\right|_{\slashed{p}=m_\ell,\, \text{IR-div}} = - \frac{\alpha_0}{2\pi}\ln\frac{m_\gamma^2}{m_\ell^2} \,.
\end{align}
Then one finds the PHOTOS contribution
%\footnote{%
%\tk{When one takes the $x_\ell \ll 1$ expansion, the PHOTOS contribution becomes
%\beq
%F^{\text{PHOTOS}} = - \frac{\alpha_0}{\pi}\left( \ln \frac{2E_{\text{max}}}{m_{D_s}} 
%    +  x_\ell \ln x_\ell + \mathcal{O}\left(x_\ell^2
%    \right)\right)\,,
%\eeq
%so the collinear divergence is canceled.
%}
%}
\begin{align}
    F^{\text{PHOTOS}} = \frac{\alpha_0}{2\pi}\left( - 2\ln \frac{2E_{\text{max}}}{m_\ell} 
    - \frac{1+x_\ell}{1-x_\ell}\ln x_\ell \right)\,,
    \label{eq:LDphotos}
\end{align}
 which should be subtracted to avoid double counting of the FSR contributions in the theoretical calculation of $\Gamma(D_s^+ \to \ell^+ \nu_\ell)^{\text{LD}}$,
 when the PHOTOS-corrected experimental results are compared to the theoretical predictions.

In experimental analyses in the BESIII,
the PHOTOS is employed without process-dependent matrix-element corrections. 
If a different PHOTOS configuration is used, the subtraction terms should be modified accordingly.

\subsection{Numerical effects on \texorpdfstring{{$|V_{cs}|$}}{|Vcs|} and CKM unitarity}
\label{seq:result}

%%%%%%%%%%%%%%%%%%%%%% 
\begin{figure}[t]
    \centering
        \includegraphics[width=0.8\linewidth]{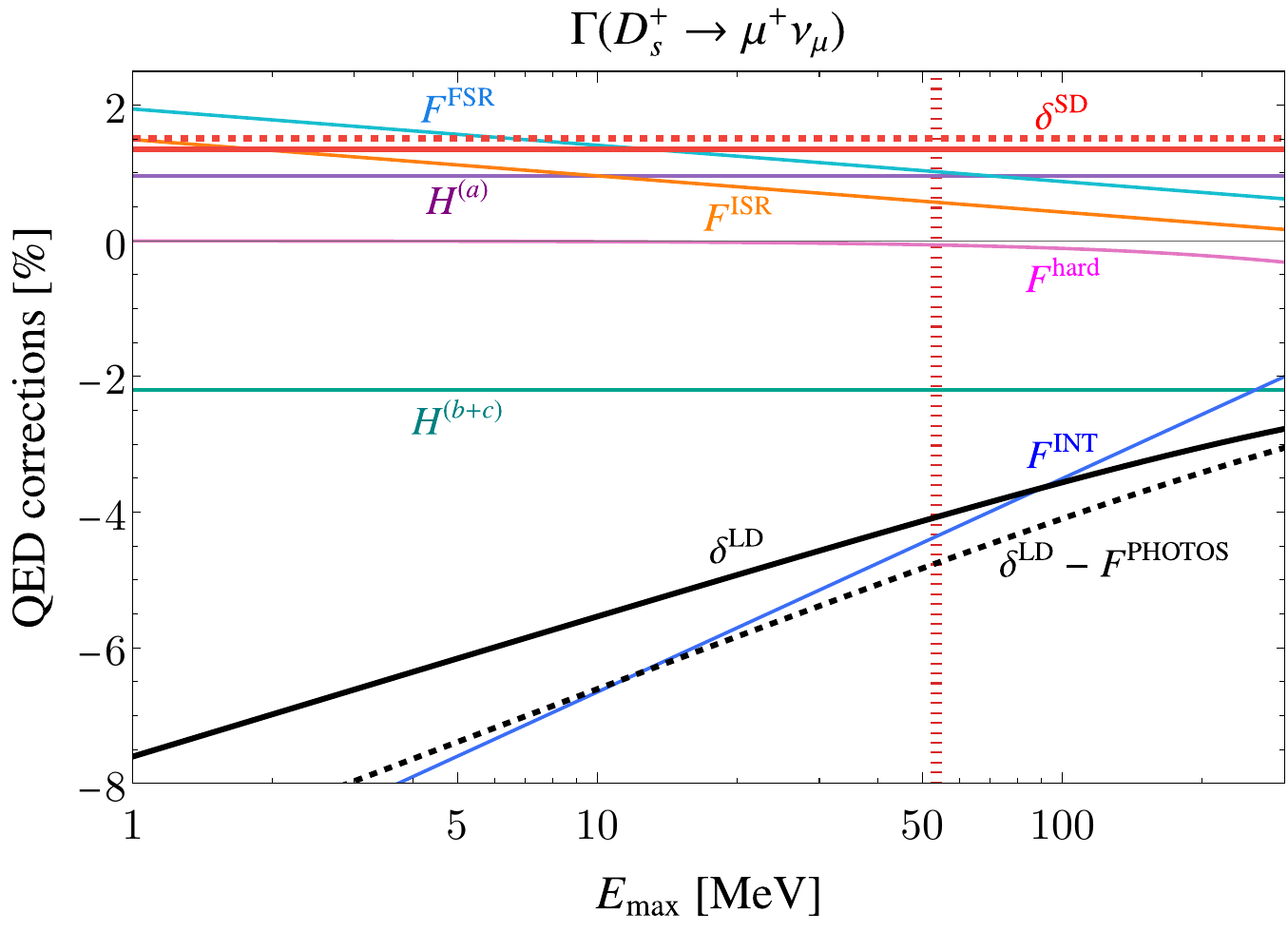}
    \\[0.5cm]    
        \includegraphics[width=0.8\linewidth]{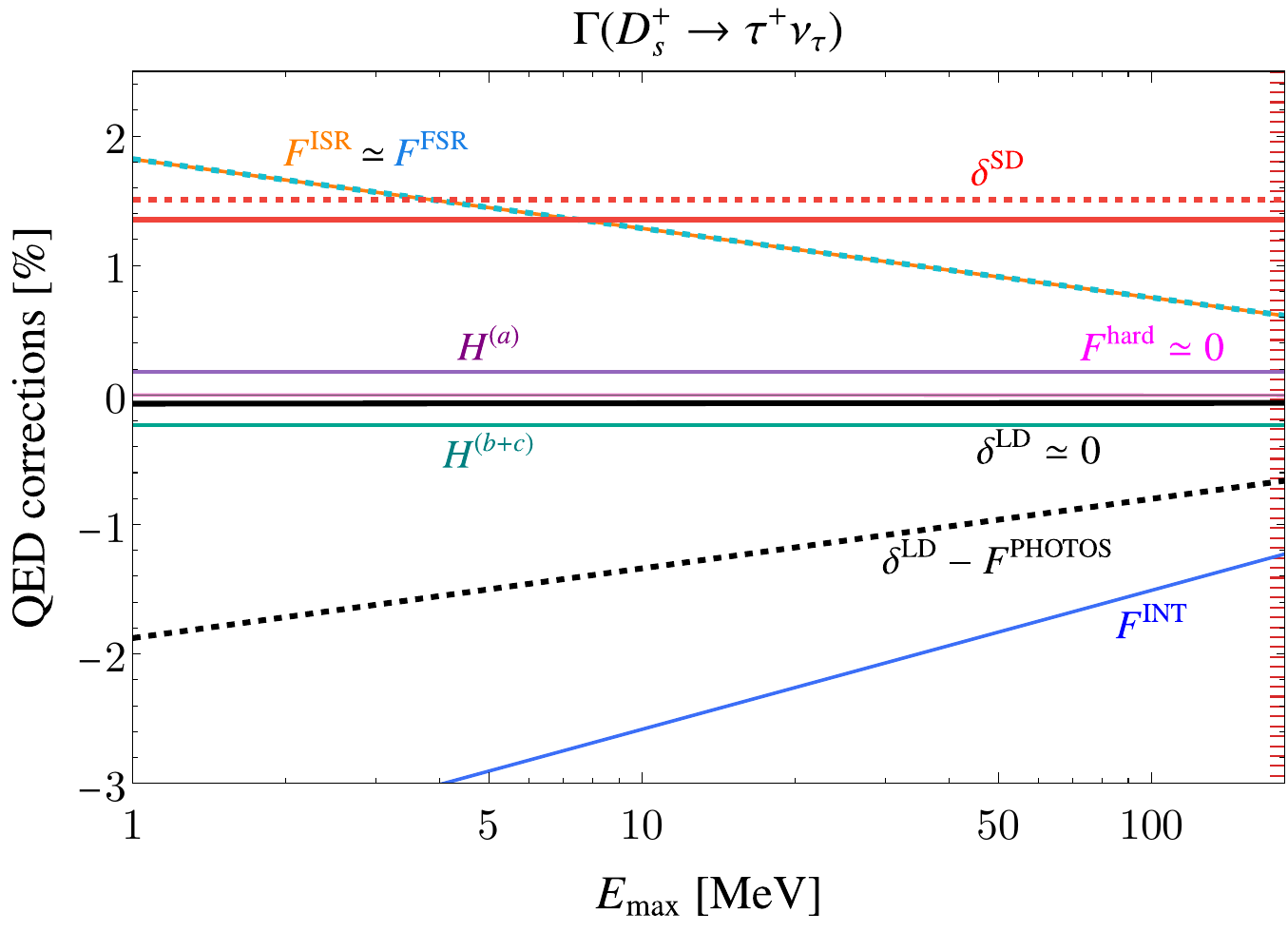}
        \caption{Each magnitude of the QED corrections to the decay rates of $D_s^+ \to \mu^+ \nu_\mu$ (upper panel) and  $D_s^+ \to \tau^+ \nu_\tau$ (lower panel).
Here, $\mu = 1$\,GeV is taken for the long-distance virtual corrections.
The short- and full long-distance corrections are shown by the red and black solid lines, respectively. 
The estimated $E_{\rm max}$ in the BESIII experiments is shown by the vertical red dotted lines.
For a detailed explanation, see the main text.
        }
        \label{fig:All}
\end{figure}
%%%%%%%%%%%%%%%%%%%%%% 
In this section, we investigate the numerical impacts of the QED corrections to $D_s^+\to\ell^+\nu$ and extract the state-of-the-art value of $|V_{cs}|$
from the latest data. 
We also assess how the test of CKM unitarity is affected by the QED corrections.

First, we compare the magnitudes of the individual QED corrections to $\Gamma(D_s^+ \to\ell^+ \nu_\ell)$ and show them in Fig.~\ref{fig:All}, as a function of $E_{\rm max}$.
The upper (lower) panel corresponds to the $\mu$ ($\tau$) mode.
The short-distance corrections are defined by $\delta^{\rm SD} = \eta_{\rm EW}^2-1$ and shown by the red-solid and red-dashed lines for the NDR and HV schemes, respectively. 
The long-distance corrections are decomposed as 
$\delta^{\rm LD} = F^{\text{ISR}} +F^{\text{FSR}}+F^{\text{INT}}+F^{\text{hard}} + H^{\text{(a)}}+H^{\text{(b}+\text{c)}}$, see Eq.~\eqref{eq:longdef}. 
For the virtual corrections $ H^{\text{(a)}}$ and $H^{\text{(b}+\text{c)}}$, we take $\mu=1$\,GeV.
The black-solid lines stand for the full long-distance QED corrections, while the black-dashed line represents the long-distance corrections after subtracting the PHOTOS QED correction in Eq.~\eqref{eq:LDphotos}.
The estimated $E_{\rm max}$ in the BESIII experiments is shown by the vertical red dotted lines. 
For $\tau$ mode, the collected data are dominated by the soft-photon inclusive signals, see discussions in Sec.~\ref{sec:Emax}.

For $\mu$ mode, it is clearly shown that all of QED corrections (except for $F^{\rm hard}$) are $\mathcal{O}(1)\%$ size and exceed the naive expected magnitude $\mathcal{O}(\alpha/\pi)\approx  0.2\%$.
The total QED correction is dominated by the interference contributions ($F^{\rm INT}$) between the ISR and FSR. Furthermore, it is partially canceled by the short-distance correction. 

For $\tau$ mode, since the emitted $\tau$ is non-relativistic, all QED corrections are suppressed than the $\mu$ mode.
Due to $m_\tau \sim m_{D_s}$ we find $F^{\rm ISR} \simeq F^{\rm FSR}$ and $F^{\rm ISR} + F^{\rm FSR} +  F^{\rm INT} \simeq 0$, and the full long-distance QED correction becomes consistent with the naive size $\mathcal{O}(\alpha/\pi)$.
However, since the simulation by the PHOTOS takes into account only the contribution from the final state radiations, 
it implies that the simulations would lead to an overestimation of the QED correction, which comes out 
as the black dashed line ($\delta^{\rm LD} - F^{\rm PHOTOS}$) in the figure.
Fortunately, we find that this overestimation is partially canceled by the short-distance correction.

Next, we extract the CKM component $|V_{cs}|$ from the latest data using the complete QED correction, subtracting the PHOTOS correction.
The master formula for the extracted $|V_{cs}|$ in the NDR scheme is
\beq
\left| V_{cs}\right|^2
&= \frac{\mathcal{B}(D_s^+\to\ell^+\nu_\ell)}{\tau_{D_s}}
\Biggl(
\frac{G_F^2}{8 \pi}  m_{D_s} m_\ell^2 f_{D_s}^2\left(1-x_\ell\right)^2
\left\{1 + 
\frac{\alpha(m_{D_s})}{2 \pi}\left[\ln \left(\frac{M_Z^2}{m_{D_s}^2}\right)+\frac{1}{2} \mathcal{A}_{g_s}-\frac{11}{6}\right]
\right\}^2 \non
&\quad  \times \Omega_B \left(\Omega_B^{\text{PHOTOS}}\right)^{-1}
\Biggl\{
1 + 
\frac{\alpha_0}{2 \pi}\left[-2 -\frac{1+x_{\ell}}{1-x_{\ell}}\left(\frac{1}{2} \ln ^2 x_{\ell}+2 \operatorname{Li}_2\left(1-x_{\ell}\right)\right)\right.\non
& \qquad \qquad \left.- \frac{3}{2} 
\ln \frac{\mu^2}{m_{D_s} m_{\ell}}
- \frac{1 + 15 x_\ell}{4 (1 - x_\ell)} \ln x_{\ell}\right]
+
F^{\text{hard}}
+\frac{\alpha_0}{2\pi}\frac{1+x_\ell}{1-x_\ell}\ln x_\ell \Biggr\}\Biggr)^{-1}\,,
\eeq
with the PHOTOS resummation factor 
\beq
\Omega_B^{\text{PHOTOS}}= \left( \frac{2 E_{\rm max}}{m_\ell}\right)^{-\frac{\alpha_0}{\pi}}\,.
\eeq
%%%%%%%%%%%%%%%%%%%%%% 
\begin{figure}[t]
\begin{center}
    \includegraphics[width=1\textwidth]{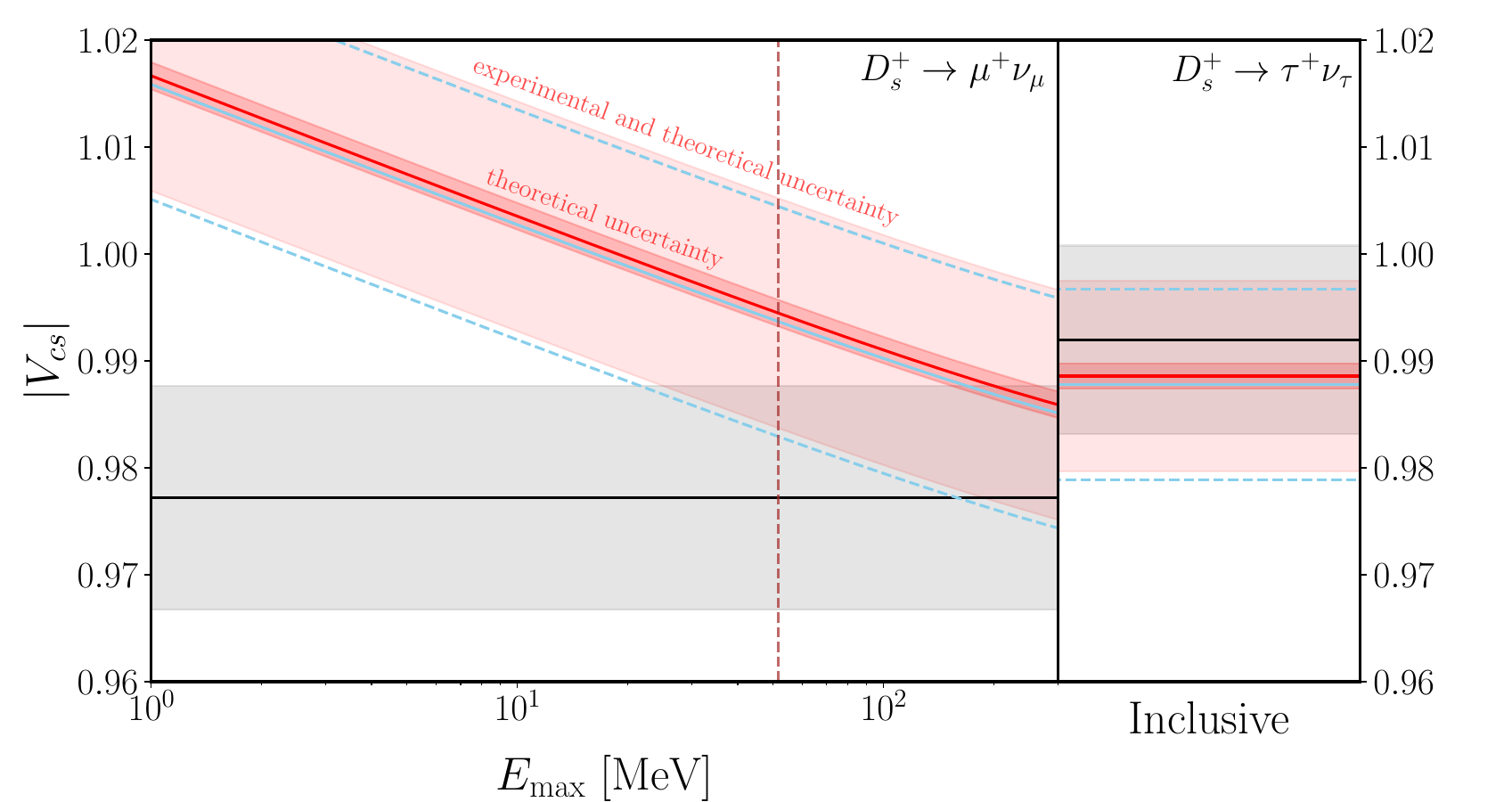}
    \caption{
    Extracted values of $|V_{cs}|$ from the latest data on $D_s^+$ leptonic decays in the $\mu$ (left panel), shown as a function of $E_{\rm max}$,  and $\tau$ channels (right panel).
    The complete QED corrections evaluated in the NDR scheme are included in the red bands,  
    while the gray bands show results obtained without any radiative corrections.
The darker bands indicate the theoretical uncertainty from the long-distance QED corrections, evaluated by varying the renormalization scale
 $500\,\text{MeV} < \mu < 2\,\text{GeV}$ (the solid lines correspond to $\mu=1\,$GeV).
The lighter bands additionally incorporate the current experimental uncertainties.
The light blue (dashed) lines show the results in the HV scheme.
}
    \label{fig:Vcs_Ds}
\end{center}
\end{figure}
%%%%%%%%%%%%%%%%%%%%%% 

For the latest data, 
we use the updated world averages from the 2025 update of the Review of Particle Physics by PDG \cite{ParticleDataGroup:2024cfk},
\beq
\begin{aligned}
\mathcal{B}\left(D_s^+ \to \mu^+\nu_\mu\right) &= \left( 5.37 \pm 0.11 \right) \times 10^{-3} \,,\\
\mathcal{B}\left(D_s^+ \to \tau^+\nu_\tau\right) &= \left( 5.39 \pm 0.09 \right) \times 10^{-2} \,,
\label{eq:BRexp}
\end{aligned}
\eeq
with $\tau_{D_s} = (5.012 \pm 0.022)\times 10^{-13}$\,s.\footnote{%
The 2024 published edition of the PDG provides 
$\mathcal{B}(D_s^+\to\mu^+\nu_\mu)=(5.35\pm 0.12)\times10^{-3}$ and $\mathcal{B}(D_s^+\to\tau^+\nu_\tau)=(5.36\pm 0.10)\times10^{-2}$ \cite{ParticleDataGroup:2024cfk}.}

The extracted values of $|V_{cs}|$ 
in each decay mode are shown in Fig.~\ref{fig:Vcs_Ds}.
The red bands include 
the complete QED corrections evaluated in the NDR scheme,  
    while the gray bands do not include any radiative corrections.
 The theoretical 
uncertainty  from the renormalization scale in the long-distance corrections is shown by the darker bands varying 
 $500\,\text{MeV} < \mu < 2\,\text{GeV}$.
 The lighter bands additionally incorporate the experimental uncertainties. 
The light blue (dashed) lines show the results in the HV scheme.  
 Hence, both the theoretical uncertainty and 
 the regularization-scheme dependence 
 are much smaller than the current sizable experimental uncertainty.
 This difference between the two schemes is
 incorporated as the theoretical uncertainty in the final result of $|V_{cs}|$ (next figure).

%%%%%%%%%%%%%%%%%%%%%% 
\begin{figure}[t]
\begin{center}
    \includegraphics[width=1.\textwidth]{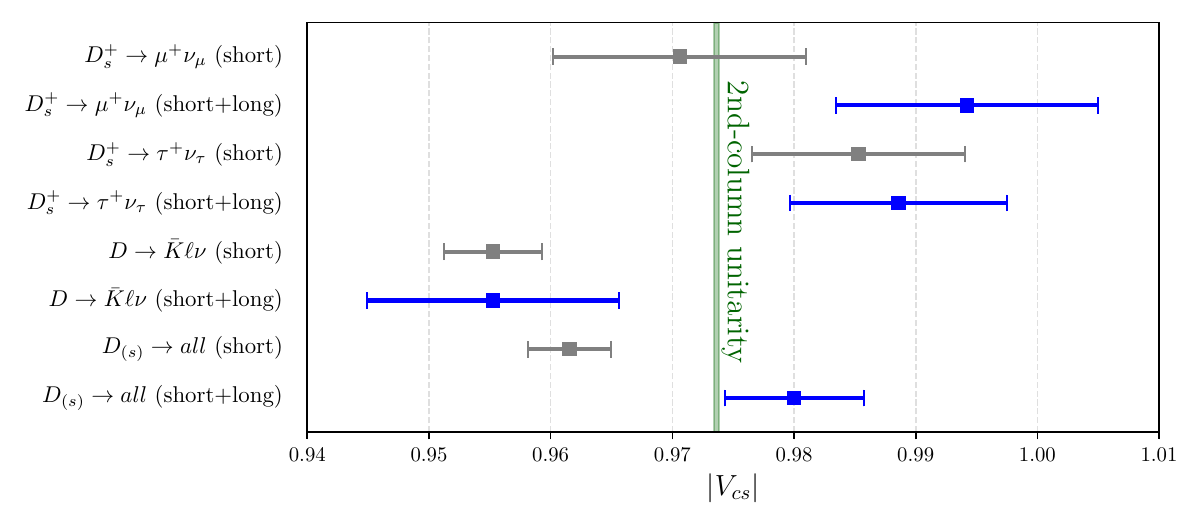}
    \caption{Summary of the determination of $|V_{cs}|$ from $D_s^+$ leptonic and $D$ semi-leptonic decays. 
    The gray bars include only short-distance corrections, while the blue bars further include the long-distance QED corrections. 
    The green band represents the prediction of the CKM unitarity test from $(V^\dagger V)_{22}$.} 
    \label{fig:Vcs}
\end{center}
\end{figure}
%%%%%%%%%%%%%%%%%%%%%%  
In addition, we evaluate the global fit of $|V_{cs}|$ incorporating the QED corrections.
It is known that the most precise determination of $|V_{cs}|$
comes from $D$-meson semi-leptonic decay measurements: $D^{0} \to {K}^- \ell^+ \nu_\ell$  and $D^{+} \to \Kb^0 \ell^+ \nu_\ell$ ($\ell = e, \mu$).
In Fig.~\ref{fig:Vcs}, we summarize the extracted values of $|V_{cs}|$ from leptonic and semi-leptonic decays 
and those fitted values. 
The $D$-meson semi-leptonic decays are carefully studied in Ref.~\cite{Bolognani:2024cmr}. 
Here, we quote the ``nominal'' scenario of Ref.~\cite{Bolognani:2024cmr}, in which 
the $D\to \Kb$ form factors utilizes the lattice results obtained by the HPQCD \cite{Parrott:2022rgu} and FNAL/MILC collaborations \cite{FermilabLattice:2022gku}, 
%then 
%we rescale it by obtaining short-distance correction in Eq.~\eqref{eq:SD-final} in the NDR scheme. 
we then rescale it by the short-distance correction obtained in Eq.~\eqref{eq:SD-final} in the NDR scheme. 
All gray bars indicate the results incorporating only short-distance corrections,
while the blue bars include the long-distance QED corrections in $D_s^+$-meson leptonic decays.
Note that calculations of 
the long-distance QED corrections in $D$-meson semi-leptonic decays are currently missing.
Following Refs.~\cite{FermilabLattice:2022gku,FlavourLatticeAveragingGroupFLAG:2024oxs}, 
we add $1\%$ uncertainty due to the missing long-distance QED corrections.
From another point of view, 
the CKM unitarity predicts the green shaded region using the PDG average values of $|V_{us}|$ and $|V_{ts}|$ \cite{ParticleDataGroup:2024cfk}.

In Ref.~\cite{Bolognani:2024cmr}, it was shown that when one considers the joint fit of $|V_{cs}|$ incorporating only short-distance corrections, then the CKM unitarity is significantly violated, which is clearly shown in this figure.  
On the other hand, 
we find that the long-distance QED corrections reduce the tension in the CKM unitarity test.
We should emphasis that this agreement of the unitarity test is partly driven by the additional systematic uncertainty assigned to the missing long-distance QED corrections in the $D$ semi-leptonic decays.

Finally, we numerically compare the fitted values of $|V_{cs}|$ with the CKM unitarity. 
First, the extracted values of $|V_{cs}|$ from $D_s^+ \to \mu^+ \nu_\mu $ and $D_s^+ \to \tau^+ \nu_\tau$ are
\beq
\left|V_{cs}\right|_{D_s\to\mu\nu} &= 0.9942 \pm 0.0020_{\rm lat.} \pm 0.0015_{\rm th.} \pm 0.0105_{\rm exp.} \,,\\
\left|V_{cs}\right|_{D_s\to\tau\nu} &= 0.9886  \pm 0.0020_{\rm lat.} \pm 0.0014_{\rm th.} \pm 0.0086_{\rm exp.} \,,
\eeq
respectively, 
where the theoretical uncertainty comes from the renormalization group scale dependence $500\,\text{MeV}<\mu < 2\,$GeV and the regularization-scheme dependence, while the experimental uncertainty is dominated by the branching ratios in Eq.~\eqref{eq:BRexp}.
Combining the two channels, we obtain 
\beq
\left|V_{cs}\right|_{D_s} = 0.991 \pm 0.007 \,.
\label{eq:VcsDsresult}
\eeq

Furthermore, combining it with the $D$ semi-leptonic decay measurements which are summarized in Table~\ref{table1}, we obtain the fitted value of $|V_{cs}|$
\beq
\left|V_{cs}\right| = 
 \begin{cases}
0.980 \pm 0.006 & \text{(nominal)}\,,\\
0.983 \pm 0.006& \text{(scale~factor)}\,,
\label{eq:finalresult}
\end{cases}
\eeq
where the ``scale factor'' scenario is defined in Ref.~\cite{Bolognani:2024cmr}: 
For the $D\to \Kb$ form factors,
in addition to the HPQCD and FNAL/MILC,
the lattice results obtained by the ETM collaboration \cite{Lubicz:2017syv,Lubicz:2018rfs}
is also used. However, it is known that the ETM results are not mutually consistent with those of HPQCD and FNAL/MILC. 
Hence, including the ETM results leads to an additional scale factor in the systematic uncertainty. As a result, the fitted result for $D$ semi-leptonic decays in the scale-factor scenario has a larger uncertainty.

\begin{table}[t]
\centering
\caption{Results of the second-column CKM unitarity test.
For $D \to \Kb \ell \nu$, only the short-distance corrections (NDR) are taken into account, while its second uncertainties come from an estimation of the size of missing long-distance QED corrections.}
\begin{tabular}{c c c c c }
\toprule
scenario & data set & $|V_{cs}|$ & $\Delta_\text{CKM}$ & \text{result}   \\
\midrule
\multirow{2}{*}{nominal} 
 & $D \to \Kb \ell \nu$ & $0.955 \pm 0.004 \pm 0.010$ & \multirow{2}{*}{$0.012\pm 0.011$} & \multirow{2}{*}{$1.1\ \sigma$}   \\
 & all average & $0.980 \pm 0.006$ &   &        \\
\midrule
\multirow{2}{*}{scale factor} 
 & $D \to \Kb \ell \nu$ & $0.959 \pm 0.007 \pm 0.010$ & \multirow{2}{*}{$0.018\pm 0.012$} & \multirow{2}{*}{$1.6\ \sigma$}   \\
 & all average & $0.983\pm0.006$ &  &    \\
\bottomrule
\end{tabular}
\label{table1}
\end{table}

Now, let us check 
 the unitarity of the CKM matrix using $|V_{cs}|$ obtained in Eq.~\eqref{eq:finalresult}.
 In Ref.~\cite{Bolognani:2024cmr},  it is reported that both the second-row and second-column unitarity conditions exhibit a significant deficit when only the short-distance corrections are taken into account.
 For the second-row unitarity test, the CKM component $|V_{cd}|$ must also be determined from data, including QED radiative corrections. 
 However, these studies are currently missing the calculation of the QED corrections \cite{FlavourLatticeAveragingGroupFLAG:2024oxs}, and $|V_{cd}|$ is also determined from the neutrino scattering data \cite{ParticleDataGroup:2024cfk}.
 On the other hand, there have been many efforts to calculate the QED corrections to $|V_{us}|$, so that we focus on the second-column CKM unitarity test,
\beq
\left(V^\dagger V\right)_{22}&=|V_{us}|^2+|V_{cs}|^2+|V_{ts}|^2=1\,.
\eeq
The measure of the CKM unitarity violation is represented by $\Delta_\text{CKM}$, defined as
\beq
\Delta_\text{CKM}\equiv |V_{us}|^2+|V_{cs}|^2+|V_{ts}|^2-1.
\eeq

By using the world averages of $|V_{us}|$ and $|V_{ts}|$ by the PDG \cite{ParticleDataGroup:2024cfk}
\beq
\left| V_{us}\right| &= 0.22431 \pm 0.00085\,,\\
\left| V_{ts}\right| &= \left(41.5 \pm 0.9\right) \times 10^{-3}\,,
\eeq
we conclude
\beq
\Delta_\text{CKM} = 
 \begin{cases}
0.012 \pm 0.011 & \text{(nominal)}\,,\\
0.018 \pm 0.012 & \text{(scale~factor)}\,,
\label{eq:finalresult2}
\end{cases}
\eeq
which are consistent with the CKM unitarity condition at  $1.1\sigma$ and $1.6\sigma$ levels, respectively.
%\ks{These are the result of being a bit more influenced by the 1\% uncertainty of the lattice.}
Furthermore, 
even if we use the result by a global fit of $|V_{us}| =  0.22405 \pm 0.00035$ in Ref.~\cite{Crivellin:2022rhw},
we obtain the same values of $\Delta_\text{CKM}$  for both scenarios.
These results are summarized in Table~\ref{table1}.
One should note that the quoted significance is sensitive to the additional uncertainty due to the missing long-distance QED corrections in  $D \to \Kb \ell \nu$.
Therefore, it should be interpreted with caution until these long-distance (and structure-dependent) QED corrections are computed explicitly.

It is found that, in the nominal scenario, the second-column CKM unitarity condition is well satisfied.
We hope that, once the long-distance QED corrections to $D \to \Kb \ell \nu$ are taken into account, the CKM unitarity can be tested more robustly
%will be more robustly checked
in future studies, as expected from Fig.~\ref{fig:Vcs}.

%=======================================================
%        CONCLUSION
%=======================================================
\section{Discussion and Conclusions}
\label{sec:conc}

In this article, we carefully studied the short-distance EW-QED corrections and the long-distance QED corrections to $D_{s}^+ \to \ell^+ \nu_\ell$ decays at one-loop level within the SM.
We analytically confirmed two minor contributions which are often neglected in previous phenomenological analyses; (1) the non-logarithmic correction in the short-distance correction and (2) the correction from the non-soft-photon radiation corresponds to $\mathcal{O}(E_{\rm max})$ contribution.
On the other hand, for the first time, the radiative corrections were calculated in both the NDR and HV schemes, and the theoretical uncertainty arising from the regularization-scheme difference was considered.

We found that properly including these radiative corrections is essential 
to bring the second-column CKM unitarity tests into agreement with the SM prediction.
Our study highlights that the present bottleneck in confirming CKM unitarity 
is the precision of the QED corrections. 
This underscores the importance of further improving lattice simulations by consistently incorporating QED effects to achieve more stringent and reliable tests of CKM unitarity.

%\ks{In this work,
%to emphasize again, because the analytically long-distance corrections to $D\to\Kb\ell^+\nu$ are not calculated; we use the estimation of the lattice alternatively, as a 1\% uncertainty. Note that the result of the CKM unitarity significance is partly dominated by this error.}

So far, since the long-distance QED corrections to $D\to \Kb\ell^+\nu$ have not yet been calculated explicitly,
we instead adopted a conservative additinal $1\%$ systematic uncertainty based on Ref.~\cite{FermilabLattice:2022gku}.
We emphasize that the resulting significance of the CKM unitarity test is partly determined by this uncertainty.

In addition, the structure-dependent QED corrections are not included.
Recently, the structure-dependent QED corrections to $B^+ \to \ell^+ \nu$ were calculated by using a gauge invariant formalism   \cite{Rowe:2024jml}.
It was found that the \emph{virtual} structure-dependent corrections are non-negligible in $B^+ \to \ell^+ \nu$, which partially cancels the long-distance QED corrections evaluated by the scalar QED.
However, both the size and sign of the virtual structure-dependent QED corrections are not clear a priori in $D_s^+ \to \ell^+ \nu$,
because the QED charges  
of the quark lines are different 
and the directions of the internal fermion lines in the loop differ 
from the $B^+$ decays. (The size is expected to be mildly suppressed compared to the ones in $B^+ \to \ell^+ \nu$ by replacement of $m_{B}\to m_{D_s}, m_b \to m_c$.)
Therefore, we leave the investigation of these effects in both $D_s^+ \to \ell^+ \nu$ and $D \to \Kb \ell^+ \nu$ to future work. 
The latter contribution is expected to be small \cite{Isidori:2020acz}.

%=======================================================
%        ACKNOWLEDGEMENTS
%=======================================================
\acknowledgments
We are profoundly grateful to Toru Goto 
for sharing his calculation notes with us, carefully checking our results, and providing many insightful comments.
We would like to thank Huijing Li, Rong-Gang Ping, and Jin Min Yang  
for valuable discussions on the $D_s^+$-meson measurements in the BESIII experiment.
We also thank Motoi Endo, Takashi Kaneko, and Roman Zwicky for useful discussions.
The work of T.K. is supported by the JSPS Grant-in-Aid for Scientific Research Grant No.\,24K22872 and 25K07276.

%=======================================================
%        APPENDIX
%=======================================================

\appendix

\section{Details of the short-distance corrections}

\subsection{Regularization scheme dependence}
\label{sec:Appreg}

We use dimensional regularization to handle the UV divergence. 
There are two regularization schemes: the naive dimensional regularization (NDR) and the 't Hooft-Veltman scheme (HV, also BMHV), which depend on how $\gamma_5$ is treated. 
One-loop four-Fermi operators have the following structures
inserted four gamma matrices in the tree-level operator:
$[\psi\Gamma \psi][\psi\Gamma\psi]\equiv \Gamma\otimes\Gamma\ $ with $\Gamma=\gamma^\nu(1-\gamma_5)/2$.
Contractions of these operators give 
 \cite{Buras:1989xd}
\begin{equation}
    \begin{aligned}        \gamma_\mu\gamma_\rho\Gamma\gamma^\rho\gamma^\mu\otimes\Gamma&=A_1(\epsilon)\Gamma\otimes\Gamma+E_1\,,\\  
    \Gamma\gamma_\mu\gamma_\rho\otimes\Gamma\gamma^\mu\gamma^\rho&=A_2(\epsilon)\Gamma\otimes\Gamma+E_2\,,\\
    \Gamma\gamma_\mu\gamma_\rho\otimes\gamma^\rho\gamma^\mu\Gamma&=A_3(\epsilon)\Gamma\otimes\Gamma+E_3\,,
    \end{aligned}
\end{equation}
where $A_i$ are given as
\begin{equation}
    \begin{aligned}
        A_1(\epsilon)=A_3(\epsilon)=
            \begin{cases}
                 4(1-\epsilon)^2\ \ \ (\rm{NDR}) \,,\\
                 4(1+\epsilon^2)\ \ \ (\rm{HV})\,,
            \end{cases}
    \end{aligned}
\end{equation}
and
\begin{equation}
    \begin{aligned}
        A_2(\epsilon)=
                 4(4-\epsilon-\epsilon^2)\ \ \ (\rm{NDR,HV})\,. \\
    \end{aligned}
\end{equation}
The extra term $E_i$ is the scheme-dependent evanescent operator.

\subsection{Matching onto the weak Hamiltonian}
\label{sec:WET}
In this section, we describe the details of the calculations of the $\gamma W$-box diagrams.
By calculating the one-loop box diagrams in Fig.~\ref{fig:box}, we obtain
\beq
       i\mathcal{M}^{\gamma W}_{\text{(a)}}
        &= i\mathcal{M}_0 Q_s Q_\ell \non 
        & \times \frac{\alpha}{\pi} \Bigg\{ \underbrace{\frac{3}{8} +\frac{1}{4}\ln \left[ \frac{2M_W^2}{s(1-\cos\theta)}\right]}_{\ell^2} - \underbrace{\frac{\pi^2}{6} - \frac{1}{4}\ln^2\left[\frac{2m_\gamma^2}{s(1-\cos\theta)}\right] - \ln\left[\frac{2m_\gamma^2}{s(1-\cos\theta)}\right] - \frac{5}{4}}_{\ell^0}  \Bigg\}\,,
        \\
        i\mathcal{M}^{\gamma W}_{\text{(d)}}
        &= i\mathcal{M}_0 Q_c Q_\ell \non  
        & \times \frac{\alpha}{\pi} \Bigg\{ \underbrace{-\frac{3}{2} -\ln \left[ \frac{2M_W^2}{s(1+\cos\theta)}\right]}_{\ell^2} + \underbrace{\frac{\pi^2}{6} + \frac{1}{4}\ln^2\left[\frac{2m_\gamma^2}{s(1+\cos\theta)}\right] + \ln\left[\frac{2m_\gamma^2}{s(1+\cos\theta)}\right]  + \frac{3}{2}}_{\ell^0}  \Bigg\}\,,
\eeq
for the terms proportional to $\ell^2$ and $\ell^0$ in the numerators of the loop calculations, respectively, 
where $\ell$ denotes the loop momentum.

By calculating the virtual $\gamma$ corrections in Fig.~\ref{fig:EFT-one-loop} to the matching onto the weak Hamiltonian, we obtain
\beq
        i\mathcal{M}^{\gamma}_{\text{(a)}}
        &= i\mathcal{M}_0 Q_s Q_\ell \non  
        & \times \frac{\alpha}{\pi} \Bigg\{ \underbrace{\frac{1}{4\bar{\epsilon}}+\frac{1}{4} +\frac{1}{4}\ln \left[ \frac{2\mu^2}{s(1-\cos\theta)}\right]}_{\ell^2} - \underbrace{\frac{\pi^2}{6} - \frac{1}{4}\ln^2\left[\frac{2m_\gamma^2}{s(1-\cos\theta)}\right] - \ln\left[\frac{2m_\gamma^2}{s(1-\cos\theta)}\right] - \frac{5}{4}}_{\ell^0}  \Bigg\}
 \,,\\
        i\mathcal{M}^{\gamma}_{\text{(b)}}
        &= i\mathcal{M}_0 Q_c Q_\ell \non  
        & \times \frac{\alpha}{\pi} \Bigg\{ \underbrace{-\frac{1}{\bar{\epsilon}}-\frac{11}{4} -\ln \left[ \frac{2\mu^2}{s(1+\cos\theta)}\right]}_{\ell^2} + \underbrace{\frac{\pi^2}{6} + \frac{1}{4}\ln^2\left[\frac{2m_\gamma^2}{s(1+\cos\theta)}\right] + \ln\left[\frac{2m_\gamma^2}{s(1+\cos\theta)}\right]  + \frac{3}{2}}_{\ell^0}  \Bigg\}.
\eeq
One can find that $l^0$ contributions from both box and four-Fermi diagrams are entirely identical in each diagram: the IR divergences, angular dependence, and $s$ dependence are totally canceled in Eqs.~\eqref{eq:cancel1} and \eqref{eq:cancel2}.

\section{Details of the long-distance corrections}

\subsection{Lagrangian}
Below the EW energy scale, the effective Lagrangian
that explains $D_s^+\to \ell^+ \nu_\ell$ is
\begin{align}
   \mathcal{L}_{c\bar{s} \to \ell^+ \nu} = - \frac{G_F}{\sqrt{2}} V_{cs}^* \left[ \overline{\nu}_\ell \gamma^\mu(1-\gamma_5) \ell \right] \left[ \overline{s} \gamma_\mu (1-\gamma_5) c  \right]\,. 
\end{align}
The tree-level decay amplitude can be obtained by
\begin{align}
    i\mathcal{M}_0
    &= -i \frac{G_F}{\sqrt{2}} V_{cs}^* \bra{\ell^+ \nu_\ell} \left[ \overline{\nu}_\ell \gamma^\mu(1-\gamma_5) \ell \right] \left[\overline{s}\gamma_\mu(1-\gamma_5) c \right] \ket{D_s^+} \non
    &= -i \frac{G_F}{\sqrt{2}} V_{cs}^* \overline{u}_{\nu_\ell} \gamma^\mu(1-\gamma_5) v_\ell \bra{0} \overline{s}\gamma_\mu(1-\gamma_5) c  \ket{D_s^+} \non
    &= i \frac{G_F}{\sqrt{2}} V_{cs}^* \overline{u}_{\nu_\ell} \gamma^\mu(1-\gamma_5) v_\ell \bra{0} \overline{s}\gamma_\mu \gamma_5 c  \ket{D_s^+} \non
    &= - \frac{G_F}{\sqrt{2}} V_{cs}^* f_{D_s} \overline{u}_{\nu_\ell} \slashed{p}_{D_s}(1-\gamma_5) v_\ell \,.
\end{align}

\subsection{Dilogarithm formula}
The following identity between four dilogarithm functions is used in the calculation of the long-distance QED corrections:
\beq
\operatorname{Li}_2\left(\frac{x}{1+x}\right)-\operatorname{Li}_2\left(\frac{1}{1+x}\right)+4 \operatorname{Li}_2\left(\frac{1-x}{1+x}\right)-\operatorname{Li}_2\left[\left(\frac{1-x}{1+x}\right)^2\right] =\ln x \ln (1+x)+2 \operatorname{Li}_2(1-x)\,,
\eeq
which is valid for $x>0$.

According to Ref.~\cite{Isidori:2007zt},
the soft-photon emission from a meson can be written as
\begin{equation}
    \begin{aligned}
        \frac{\Gamma(M\to \ell \nu \gamma_{\text{soft}})}{\Gamma(M\to \ell \nu)} = 
        \frac{\alpha}{2\pi} \sum_{i,j=0}^{1} Q_i Q_j \left[ 4b_{ij}\ln \left( \frac{m_\gamma}{2E_{\text{max}}}\right) + 2F_{ij} + \mathcal{O}(E_{\text{max}}) \right]\,,
    \end{aligned}
\end{equation}
where 
$
Q_{1} = Q_\ell
$
and  $Q_0=-Q_\ell
$.
The functions of $b_{ij}$ and $F_{ij}$ are defined in Ref.~\cite{Isidori:2007zt}.

The coefficient of the $\ln m_\gamma$ term can be expressed as
\begin{equation}
    \begin{aligned}
        4\left(b_{00} + b_{11} - 2b_{01}\right) 
        = 4 \left( 1 + \frac{1+x_\ell}{2(1-x_\ell)} \ln x_\ell \right)   \,,
    \end{aligned}
\end{equation}
where $x_\ell = m_\ell^2/m_{D_s}^2$. 
This coefficient is consistent with our result in Eq.~\eqref{eq:result_photon_emission}.

Furthermore, the finite corrections $F_{ij}$ can be expressed as
\beq
        2\left(F_{00} + F_{11} - 2F_{01}\right) 
        & = 2 - \frac{1+x_\ell}{1-x_\ell} \ln x_\ell - \frac{1+x_\ell}{1-x_\ell} \Biggl\{ \frac{1}{2} \ln^2 x_\ell - \ln x_\ell \ln(1+x_\ell) \non
        &~
        + \operatorname{Li}_2\left(\frac{x_\ell}{1+x_\ell} \right) 
        - \operatorname{Li}_2\left(\frac{1}{1+x_\ell} \right)  
        + 4\operatorname{Li}_2\left(\frac{1-x_\ell}{1+x_\ell} \right)
        - \operatorname{Li}_2\left[\left(\frac{1-x_\ell}{1+x_\ell} \right)^2\right]  \Biggr\} 
         \non
        &= 2 - \frac{1+x_\ell}{1-x_\ell} \left[\ln x_\ell +  \frac{1}{2} \ln^2 x_\ell + 2\operatorname{Li}_2(1-x_\ell)
        \right]  \,,
\eeq
which is again consistent with our result in Eq.~\eqref{eq:result_photon_emission}.

\subsection{Photon energy cut}
\label{sec:Dalitz}
%%%%%%%%%%%%%%%%%%%%%% 
\begin{figure}[t]
\begin{center}
    \includegraphics[width=0.8\textwidth]{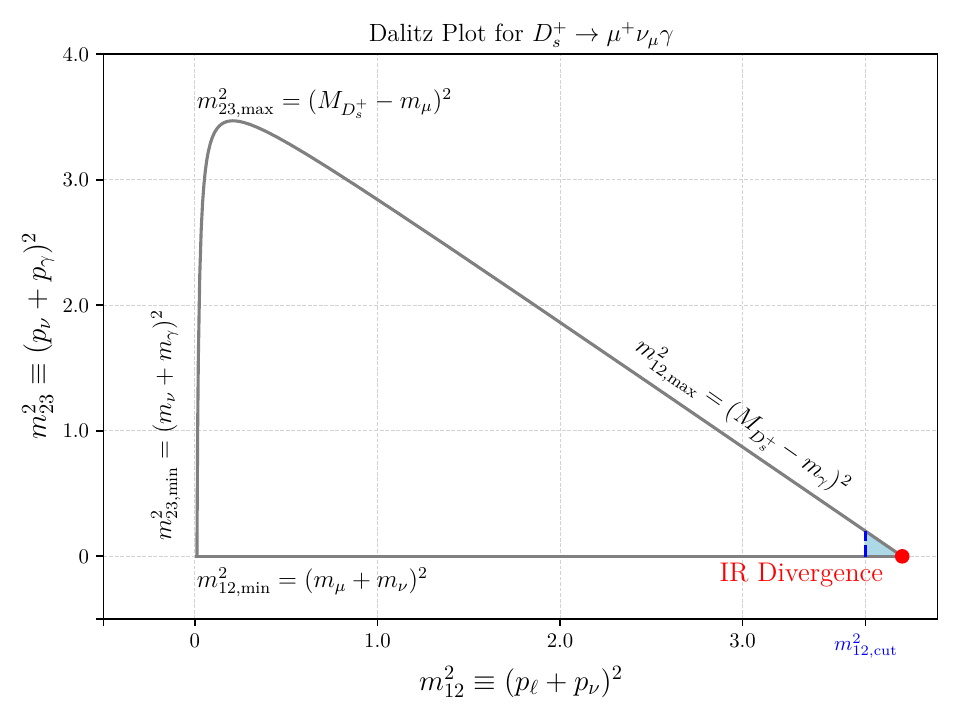}
    \caption{
        The Dalitz plot for $D_s^+\to \mu \nu_\mu \gamma$ is shown in units of [GeV$^2$].
        The entire region enclosed by the gray boundary corresponds to the inclusive limit, while the blue-shaded area represents the soft-photon region relevant to the present study. 
        The cutoff value is given by
        $m_{12,\mathrm{cut}}^2 =(p_\ell+p_\nu)^2=(p_{D_s}-p_\gamma)^2= m_{D_s}^2 - 2 m_{D_s} E_{\mathrm{max}}$,
        and the red point denotes the position at which the IR divergence appears, where the calculation has been performed using the soft-photon approximation in Eq.~\eqref{eq:softapp}. 
    } 
    \label{fig:dalitz}
\end{center}
\end{figure}
%%%%%%%%%%%%%%%%%%%%%% 
Soft photon approximation is valid for small photon energies, and if possible, it is desired not to use the approximation. Therefore, we introduced the following two photon energy cuts: 
\beq
    \begin{aligned}
        0 & \leq E_\gamma < E_{\text{cut}}  \qquad &(\text{for IR limit})\,, \\
        E_{\text{cut}} & \leq E_\gamma \leq E_{\text{max}}  \qquad &(\text{for 3-body decay})\,.
    \end{aligned}
\eeq
The IR limit is shown by the red point in Fig.~\ref{fig:dalitz},
where we used the soft photon approximation and took the $E_{\text{cut}} \to 0$. 
In the remaining shaded area, 
we used the Dalitz plot integration. 
When one integrates the entire region surrounded by the gray line, it represents the inclusive limit. 

\subsection{Inclusive limit}
\label{sec:inclusive}
%%%%%%%%%%%%%%%%%%%%%% 
\begin{figure}[t]
\begin{center}
    \includegraphics[width=0.8\textwidth]{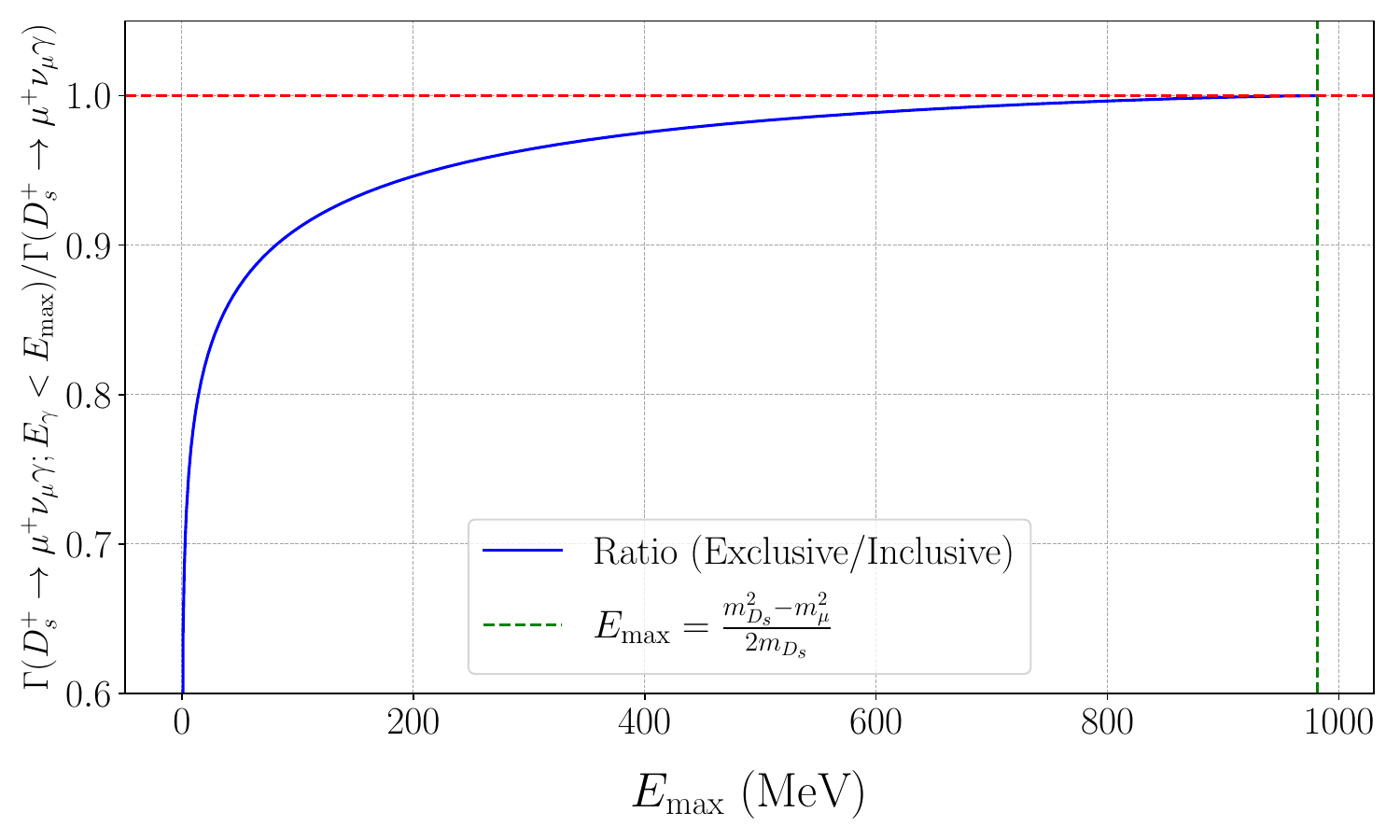}
    \caption{The ratio of the single photon radiation with $
E_\gamma<E_{\rm max}
$ in Eq.~\eqref{eq:result_photon_emission} to the inclusive one in Eq.~\eqref{eq:inclusive} is  indicated by the blue line.
It equals to one at the $E_{\text{max}} \to (m_{D_s}^2 - m_\mu^2)/2m_{D_s}$.
    }
    \label{fig:inclusive_exclusive}
\end{center}
\end{figure}
%%%%%%%%%%%%%%%%%%%%%% 
The inner-bremsstrahlung ratio for the inclusive $D_s^+$ leptonic decay is
\begin{equation}
    \begin{aligned}
        \frac{\Delta P_{\text{IB}}}{P_0}
        = \frac{\alpha}{\pi}&
        \Bigg[ 2 \left( 1 + \frac{1+x_\ell}{2(1-x_\ell)} \ln x_\ell \right) \left(\ln \frac{m_\gamma}{m_{D_s}} - \ln(1-x_\ell) - \frac{1}{4} \ln x_\ell + \frac{3}{4} \right) \\
         &\quad - \frac{x_\ell(10-7x_\ell)}{4(1-x_\ell)^2} \ln x_\ell -\frac{2(1+x_\ell)}{1-x_\ell} \operatorname{Li}_2(1-x_\ell) + \frac{15-21x_\ell}{8(1-x_\ell)} \Bigg]\,,
         \label{eq:inclusive}
    \end{aligned}
\end{equation}
which is consistent with Eq.~(4) of Ref.~\cite{Kinoshita:1959ha}.
We numerically checked it
and this is readily verified by plotting in Fig.~\ref{fig:inclusive_exclusive},
where the IR divergence is completely canceled by the following virtual corrections.
The virtual photon vertex correction is 
\begin{equation}
    \begin{aligned}
        \frac{\Delta \Gamma_{\text{vert}}}{\Gamma_0}
        = \frac{\alpha}{2\pi}\left[ -2 \left( \frac{1}{\bar{\epsilon}} + \ln \frac{\mu^2 }{m_{D_s}^2}\right) + \frac{1-3x_\ell}{1-x_\ell} \ln x_\ell -2 - 2\left( \frac{1+x_\ell}{1-x_\ell}\right) \ln x_\ell \ln \frac{m_\gamma}{\sqrt{m_{D_s}m_\ell}} \right]\,,
    \end{aligned}
\end{equation}
and the self-energy contribution is 
\begin{equation}
    \begin{aligned}
        \frac{\Delta \Gamma_{\text{self}}}{\Gamma_0} =
        \frac{\alpha}{2\pi}
        \left[ \frac{1}{2}\left( \frac{1}{\bar{\epsilon}}+ \ln \frac{\mu^2}{m_{D_s}^2}\right)+ \frac{1}{2} \ln \frac{m_\ell^2}{m_{D_s}^2} - 2 - 4 \ln \frac{m_\gamma}{\sqrt{m_{D_s}m_\ell}}\right]\,.
    \end{aligned}
\end{equation}

%\newpage
%=======================================================
%        REFERENCES
%=======================================================
\bibliographystyle{utphys28mod}
\bibliography{ref}

\providecommand{\href}[2]{#2}\begingroup\raggedright\begin{thebibliography}{10}

\bibitem{Kobayashi:1973fv}
M.~Kobayashi and T.~Maskawa, ``{CP Violation in the Renormalizable Theory of Weak Interaction},'' \href{https://dx.doi.org/10.1143/PTP.49.652}{Prog.\  Theor.\  Phys.\  {\bfseries 49} (1973) 652--657}.

\bibitem{Belfatto:2019swo}
B.~Belfatto, R.~Beradze, and Z.~Berezhiani, ``{The CKM unitarity problem: A trace of new physics at the TeV scale?}'' \href{https://dx.doi.org/10.1140/epjc/s10052-020-7691-6}{Eur.\  Phys.\  J.\  C {\bfseries 80} (2020) 149} {\ttfamily [\href{https://arxiv.org/abs/1906.02714}{arXiv:1906.02714}]}.

\bibitem{Cheung:2020vqm}
K.~Cheung, W.-Y.~Keung, C.-T.~Lu, and P.-Y.~Tseng, ``{Vector-like Quark Interpretation for the CKM Unitarity Violation, Excess in Higgs Signal Strength, and Bottom Quark Forward-Backward Asymmetry},'' \href{https://dx.doi.org/10.1007/JHEP05(2020)117}{JHEP {\bfseries 05} (2020) 117} {\ttfamily [\href{https://arxiv.org/abs/2001.02853}{arXiv:2001.02853}]}.

\bibitem{Belfatto:2021jhf}
B.~Belfatto and Z.~Berezhiani, ``{Are the CKM anomalies induced by vector-like quarks? Limits from flavor changing and Standard Model precision tests},'' \href{https://dx.doi.org/10.1007/JHEP10(2021)079}{JHEP {\bfseries 10} (2021) 079} {\ttfamily [\href{https://arxiv.org/abs/2103.05549}{arXiv:2103.05549}]}.

\bibitem{Branco:2021vhs}
G.~C.~Branco, J.~T.~Penedo, P.~M.~F.~Pereira, M.~N.~Rebelo, and J.~I.~Silva-Marcos, ``{Addressing the CKM unitarity problem with a vector-like up quark},'' \href{https://dx.doi.org/10.1007/JHEP07(2021)099}{JHEP {\bfseries 07} (2021) 099} {\ttfamily [\href{https://arxiv.org/abs/2103.13409}{arXiv:2103.13409}]}.

\bibitem{Crivellin:2022rhw}
A.~Crivellin, M.~Kirk, T.~Kitahara, and F.~Mescia, ``{Global fit of modified quark couplings to EW gauge bosons and vector-like quarks in light of the Cabibbo angle anomaly},'' \href{https://dx.doi.org/10.1007/JHEP03(2023)234}{JHEP {\bfseries 03} (2023) 234} {\ttfamily [\href{https://arxiv.org/abs/2212.06862}{arXiv:2212.06862}]}.

\bibitem{Endo:2020tkb}
M.~Endo and S.~Mishima, ``{Muon $g-2$ and CKM unitarity in extra lepton models},'' \href{https://dx.doi.org/10.1007/JHEP08(2020)004}{JHEP {\bfseries 08} (2020) 004} {\ttfamily [\href{https://arxiv.org/abs/2005.03933}{arXiv:2005.03933}]}.

\bibitem{Crivellin:2020ebi}
A.~Crivellin, F.~Kirk, C.~A.~Manzari, and M.~Montull, ``{Global Electroweak Fit and Vector-Like Leptons in Light of the Cabibbo Angle Anomaly},'' \href{https://dx.doi.org/10.1007/JHEP12(2020)166}{JHEP {\bfseries 12} (2020) 166} {\ttfamily [\href{https://arxiv.org/abs/2008.01113}{arXiv:2008.01113}]}.

\bibitem{Kirk:2020wdk}
M.~Kirk, ``{Cabibbo anomaly versus electroweak precision tests: An exploration of extensions of the Standard Model},'' \href{https://dx.doi.org/10.1103/PhysRevD.103.035004}{Phys.\  Rev.\  D {\bfseries 103} (2021) 035004} {\ttfamily [\href{https://arxiv.org/abs/2008.03261}{arXiv:2008.03261}]}.

\bibitem{Hocker:2001xe}
A.~Hocker, H.~Lacker, S.~Laplace, and F.~Le~Diberder, ``{A New approach to a global fit of the CKM matrix},'' \href{https://dx.doi.org/10.1007/s100520100729}{Eur.\  Phys.\  J.\  C {\bfseries 21} (2001) 225--259} {\ttfamily [\href{https://arxiv.org/abs/hep-ph/0104062}{hep-ph/0104062}]}.

\bibitem{Charles:2004jd}
{\bfseries CKMfitter Group} Collaboration, ``{CP violation and the CKM matrix: Assessing the impact of the asymmetric $B$ factories},'' \href{https://dx.doi.org/10.1140/epjc/s2005-02169-1}{Eur.\  Phys.\  J.\  C {\bfseries 41} (2005) 1--131} {\ttfamily [\href{https://arxiv.org/abs/hep-ph/0406184}{hep-ph/0406184}]}.

\bibitem{UTfit:2005ras}
{\bfseries UTfit} Collaboration, ``{The 2004 UTfit collaboration report on the status of the unitarity triangle in the standard model},'' \href{https://dx.doi.org/10.1088/1126-6708/2005/07/028}{JHEP {\bfseries 07} (2005) 028} {\ttfamily [\href{https://arxiv.org/abs/hep-ph/0501199}{hep-ph/0501199}]}.

\bibitem{UTfit:2007eik}
{\bfseries UTfit} Collaboration, ``{Model-independent constraints on $\Delta F=2$ operators and the scale of new physics},'' \href{https://dx.doi.org/10.1088/1126-6708/2008/03/049}{JHEP {\bfseries 03} (2008) 049} {\ttfamily [\href{https://arxiv.org/abs/0707.0636}{arXiv:0707.0636}]}.

\bibitem{Buchalla:1996fp}
G.~Buchalla and A.~J.~Buras, ``{$K \to \pi \nu \bar \nu$ and high precision determinations of the CKM matrix},'' \href{https://dx.doi.org/10.1103/PhysRevD.54.6782}{Phys.\  Rev.\  D {\bfseries 54} (1996) 6782--6789} {\ttfamily [\href{https://arxiv.org/abs/hep-ph/9607447}{hep-ph/9607447}]}.

\bibitem{Buras:2006gb}
A.~J.~Buras, M.~Gorbahn, U.~Haisch, and U.~Nierste, ``{Charm quark contribution to $K^+ \to \pi^+ \nu \bar\nu$ at next-to-next-to-leading order},'' \href{https://dx.doi.org/10.1007/JHEP11(2012)167}{JHEP {\bfseries 11} (2006) 002} {\ttfamily [\href{https://arxiv.org/abs/hep-ph/0603079}{hep-ph/0603079}]}. [Erratum: JHEP 11, 167 (2012)].

\bibitem{Lehner:2015jga}
C.~Lehner, E.~Lunghi, and A.~Soni, ``{Emerging lattice approach to the K-Unitarity Triangle},'' \href{https://dx.doi.org/10.1016/j.physletb.2016.04.064}{Phys.\  Lett.\  B {\bfseries 759} (2016) 82--90} {\ttfamily [\href{https://arxiv.org/abs/1508.01801}{arXiv:1508.01801}]}.

\bibitem{Lunghi:2024sjy}
E.~Lunghi and A.~Soni, ``{Light quark loops in $ {K}^{\pm}\to {\pi}^{\pm}\nu \overline{\nu} $ from vector meson dominance and update on the Kaon Unitarity Triangle},'' \href{https://dx.doi.org/10.1007/JHEP12(2024)097}{JHEP {\bfseries 12} (2024) 097} {\ttfamily [\href{https://arxiv.org/abs/2408.11190}{arXiv:2408.11190}]}.

\bibitem{Dery:2025pcx}
A.~Dery, ``{Natural complex plane for kaon CKM data: Framework, status, and future prospects},'' \href{https://dx.doi.org/10.1103/dz6j-q85l}{Phys.\  Rev.\  D {\bfseries 112} (2025) 053005} {\ttfamily [\href{https://arxiv.org/abs/2504.12386}{arXiv:2504.12386}]}.

\bibitem{FlavourLatticeAveragingGroupFLAG:2024oxs}
{\bfseries Flavour Lattice Averaging Group (FLAG)} Collaboration, ``{FLAG Review 2024}.'' {\ttfamily \href{https://arxiv.org/abs/2411.04268}{arXiv:2411.04268}}.

\bibitem{Yue:2013qrc}
A.~T.~Yue, {\em et al.}, ``{Improved Determination of the Neutron Lifetime},'' \href{https://dx.doi.org/10.1103/PhysRevLett.111.222501}{Phys.\  Rev.\  Lett.\  {\bfseries 111} (2013) 222501} {\ttfamily [\href{https://arxiv.org/abs/1309.2623}{arXiv:1309.2623}]}.

\bibitem{Kitahara:2023xab}
T.~Kitahara and K.~Tobioka, ``{MeV sterile neutrino in light of the Cabibbo-angle anomaly},'' \href{https://dx.doi.org/10.1103/PhysRevD.108.115034}{Phys.\  Rev.\  D {\bfseries 108} (2023) 115034} {\ttfamily [\href{https://arxiv.org/abs/2308.13003}{arXiv:2308.13003}]}.

\bibitem{Ma:2023kfr}
P.-X.~Ma, {\em et al.}, ``{Lattice QCD Calculation of Electroweak Box Contributions to Superallowed Nuclear and Neutron Beta Decays},'' \href{https://dx.doi.org/10.1103/PhysRevLett.132.191901}{Phys.\  Rev.\  Lett.\  {\bfseries 132} (2024) 191901} {\ttfamily [\href{https://arxiv.org/abs/2308.16755}{arXiv:2308.16755}]}.

\bibitem{Moretti:2025qxt}
F.~Moretti, M.~Gorbahn, and S.~J{\"a}ger, ``{Beyond Leading Logarithms in $g_V$: The Semileptonic Weak Hamiltonian at $\mathcal{O}(\alpha\,\alpha_s^2)$}.'' {\ttfamily \href{https://arxiv.org/abs/2510.27648}{arXiv:2510.27648}}.

\bibitem{Cao:2025zxs}
Z.~Cao, R.~J.~Hill, R.~Plestid, and P.~Vander~Griend, ``{The $Z\alpha^2$ correction to superallowed beta decays in effective field theory and implications for $|V_{ud}|$}.'' {\ttfamily \href{https://arxiv.org/abs/2511.05446}{arXiv:2511.05446}}.

\bibitem{Crosas:2025xyv}
{\`O}.~L.~Crosas and E.~Mereghetti, ``{Radiative corrections to superallowed beta decays at $\mathcal{O}(\alpha^2 Z)$}.'' {\ttfamily \href{https://arxiv.org/abs/2511.05481}{arXiv:2511.05481}}.

\bibitem{Seng:2022epj}
C.-Y.~Seng and M.~Gorchtein, ``{Electroweak nuclear radii constrain the isospin breaking correction to Vud},'' \href{https://dx.doi.org/10.1016/j.physletb.2022.137654}{Phys.\  Lett.\  B {\bfseries 838} (2023) 137654} {\ttfamily [\href{https://arxiv.org/abs/2208.03037}{arXiv:2208.03037}]}.

\bibitem{Seng:2023cgl}
C.-Y.~Seng and M.~Gorchtein, ``{Data-driven reevaluation of ft values in superallowed {\ensuremath{\beta}} decays},'' \href{https://dx.doi.org/10.1103/PhysRevC.109.045501}{Phys.\  Rev.\  C {\bfseries 109} (2024) 045501} {\ttfamily [\href{https://arxiv.org/abs/2309.16893}{arXiv:2309.16893}]}.

\bibitem{Gorchtein:2025wli}
M.~Gorchtein, V.~Katyal, B.~Ohayon, B.~K.~Sahoo, and C.-Y.~Seng, ``{Cabibbo-Kobayashi-Maskawa unitarity deficit reduction via finite nuclear size},'' \href{https://dx.doi.org/10.1103/z8g6-9j25}{Phys.\  Rev.\  Res.\  {\bfseries 7} (2025) L042002} {\ttfamily [\href{https://arxiv.org/abs/2502.17070}{arXiv:2502.17070}]}.

\bibitem{Grossman:2019bzp}
Y.~Grossman, E.~Passemar, and S.~Schacht, ``{On the Statistical Treatment of the Cabibbo Angle Anomaly},'' \href{https://dx.doi.org/10.1007/JHEP07(2020)068}{JHEP {\bfseries 07} (2020) 068} {\ttfamily [\href{https://arxiv.org/abs/1911.07821}{arXiv:1911.07821}]}.

\bibitem{Coutinho:2019aiy}
A.~M.~Coutinho, A.~Crivellin, and C.~A.~Manzari, ``{Global Fit to Modified Neutrino Couplings and the Cabibbo-Angle Anomaly},'' \href{https://dx.doi.org/10.1103/PhysRevLett.125.071802}{Phys.\  Rev.\  Lett.\  {\bfseries 125} (2020) 071802} {\ttfamily [\href{https://arxiv.org/abs/1912.08823}{arXiv:1912.08823}]}.

\bibitem{Carrasco:2015xwa}
N.~Carrasco, {\em et al.}, ``{QED Corrections to Hadronic Processes in Lattice QCD},'' \href{https://dx.doi.org/10.1103/PhysRevD.91.074506}{Phys.\  Rev.\  D {\bfseries 91} (2015) 074506} {\ttfamily [\href{https://arxiv.org/abs/1502.00257}{arXiv:1502.00257}]}.

\bibitem{Lubicz:2016xro}
V.~Lubicz, {\em et al.}, ``{Finite-Volume QED Corrections to Decay Amplitudes in Lattice QCD},'' \href{https://dx.doi.org/10.1103/PhysRevD.95.034504}{Phys.\  Rev.\  D {\bfseries 95} (2017) 034504} {\ttfamily [\href{https://arxiv.org/abs/1611.08497}{arXiv:1611.08497}]}.

\bibitem{Giusti:2017dwk}
D.~Giusti, {\em et al.}, ``{First lattice calculation of the QED corrections to leptonic decay rates},'' \href{https://dx.doi.org/10.1103/PhysRevLett.120.072001}{Phys.\  Rev.\  Lett.\  {\bfseries 120} (2018) 072001} {\ttfamily [\href{https://arxiv.org/abs/1711.06537}{arXiv:1711.06537}]}.

\bibitem{DiCarlo:2019thl}
M.~Di~Carlo, {\em et al.}, ``{Light-meson leptonic decay rates in lattice QCD+QED},'' \href{https://dx.doi.org/10.1103/PhysRevD.100.034514}{Phys.\  Rev.\  D {\bfseries 100} (2019) 034514} {\ttfamily [\href{https://arxiv.org/abs/1904.08731}{arXiv:1904.08731}]}.

\bibitem{Frezzotti:2020bfa}
R.~Frezzotti, {\em et al.}, ``{Comparison of lattice QCD+QED predictions for radiative leptonic decays of light mesons with experimental data},'' \href{https://dx.doi.org/10.1103/PhysRevD.103.053005}{Phys.\  Rev.\  D {\bfseries 103} (2021) 053005} {\ttfamily [\href{https://arxiv.org/abs/2012.02120}{arXiv:2012.02120}]}.

\bibitem{Bolognani:2024cmr}
C.~Bolognani, M.~Reboud, D.~van Dyk, and K.~K.~Vos, ``{Constraining |V$_{cs}$| and physics beyond the Standard Model from exclusive (semi)leptonic charm decays},'' \href{https://dx.doi.org/10.1007/JHEP09(2024)099}{JHEP {\bfseries 09} (2024) 099} {\ttfamily [\href{https://arxiv.org/abs/2407.06145}{arXiv:2407.06145}]}.

\bibitem{Wolfenstein:1983yz}
L.~Wolfenstein, ``{Parametrization of the Kobayashi-Maskawa Matrix},'' \href{https://dx.doi.org/10.1103/PhysRevLett.51.1945}{Phys.\  Rev.\  Lett.\  {\bfseries 51} (1983) 1945}.

\bibitem{deBoer:2018ipi}
S.~de~Boer, T.~Kitahara, and I.~Nisandzic, ``{Soft-Photon Corrections to $\bar{B} \to D \tau^{-} \bar{\nu}_{\tau}$ Relative to $\bar{B} \to D \mu^{-} \bar{\nu}_{\mu}$},'' \href{https://dx.doi.org/10.1103/PhysRevLett.120.261804}{Phys.\  Rev.\  Lett.\  {\bfseries 120} (2018) 261804} {\ttfamily [\href{https://arxiv.org/abs/1803.05881}{arXiv:1803.05881}]}.

\bibitem{Buras:1998raa}
A.~J.~Buras, ``{Weak Hamiltonian, CP violation and rare decays},'' in {\em {Les Houches Summer School in Theoretical Physics, Session 68: Probing the Standard Model of Particle Interactions}}, pp.~281--539.
\newblock 1998.
\newblock {\ttfamily \href{https://arxiv.org/abs/hep-ph/9806471}{hep-ph/9806471}}.

\bibitem{Sirlin:1977sv}
A.~Sirlin, ``{Current Algebra Formulation of Radiative Corrections in Gauge Theories and the Universality of the Weak Interactions},'' \href{https://dx.doi.org/10.1103/RevModPhys.50.573}{Rev.\  Mod.\  Phys.\  {\bfseries 50} (1978) 573}. [Erratum: Rev.Mod.Phys. 50, 905 (1978)].

\bibitem{Sirlin:1981ie}
A.~Sirlin, ``{Large $m_W, m_Z$ Behavior of the $O(\alpha)$ Corrections to Semileptonic Processes Mediated by $W$},'' \href{https://dx.doi.org/10.1016/0550-3213(82)90303-0}{Nucl.\  Phys.\  B {\bfseries 196} (1982) 83--92}.

\bibitem{Marciano:1983ss}
W.~J.~Marciano and A.~Sirlin, ``{On Some General Properties of the $O(\alpha)$ Corrections to Parity Violation in Atoms},'' \href{https://dx.doi.org/10.1103/PhysRevD.29.75}{Phys.\  Rev.\  D {\bfseries 29} (1984) 75}. [Erratum: Phys.Rev.D 31, 213 (1985)].

\bibitem{Marciano:1985pd}
W.~J.~Marciano and A.~Sirlin, ``{Radiative Corrections to $\beta$ Decay and the Possibility of a Fourth Generation},'' \href{https://dx.doi.org/10.1103/PhysRevLett.56.22}{Phys.\  Rev.\  Lett.\  {\bfseries 56} (1986) 22}.

\bibitem{ParticleDataGroup:2024cfk}
{\bfseries Particle Data Group} Collaboration, ``{Review of particle physics},'' \href{https://dx.doi.org/10.1103/PhysRevD.110.030001}{Phys.\  Rev.\  D {\bfseries 110} (2024) 030001}. {and 2025 update available at the PDG website \url{https://pdg.lbl.gov}}.

\bibitem{Erler:1998sy}
J.~Erler, ``{Calculation of the QED coupling $\hat \alpha (M_Z)$ in the modified minimal subtraction scheme},'' \href{https://dx.doi.org/10.1103/PhysRevD.59.054008}{Phys.\  Rev.\  D {\bfseries 59} (1999) 054008} {\ttfamily [\href{https://arxiv.org/abs/hep-ph/9803453}{hep-ph/9803453}]}.

\bibitem{Sirlin:1980nh}
A.~Sirlin, ``{Radiative Corrections in the $SU(2)_L \times U(1)$ Theory: A Simple Renormalization Framework},'' \href{https://dx.doi.org/10.1103/PhysRevD.22.971}{Phys.\  Rev.\  D {\bfseries 22} (1980) 971--981}.

\bibitem{Stuart:1991cc}
R.~G.~Stuart, ``{General renormalization of the gauge invariant perturbation expansion near the Z0 resonance},'' \href{https://dx.doi.org/10.1016/0370-2693(91)91842-J}{Phys.\  Lett.\  B {\bfseries 272} (1991) 353--358}.

\bibitem{vanRitbergen:1999fi}
T.~van Ritbergen and R.~G.~Stuart, ``{On the precise determination of the Fermi coupling constant from the muon lifetime},'' \href{https://dx.doi.org/10.1016/S0550-3213(99)00572-6}{Nucl.\  Phys.\  B {\bfseries 564} (2000) 343--390} {\ttfamily [\href{https://arxiv.org/abs/hep-ph/9904240}{hep-ph/9904240}]}.

\bibitem{Buras:1989xd}
A.~J.~Buras and P.~H.~Weisz, ``{QCD Nonleading Corrections to Weak Decays in Dimensional Regularization and 't Hooft-Veltman Schemes},'' \href{https://dx.doi.org/10.1016/0550-3213(90)90223-Z}{Nucl.\  Phys.\  B {\bfseries 333} (1990) 66--99}.

\bibitem{tHooft:1972tcz}
G.~'t~Hooft and M.~J.~G.~Veltman, ``{Regularization and Renormalization of Gauge Fields},'' \href{https://dx.doi.org/10.1016/0550-3213(72)90279-9}{Nucl.\  Phys.\  B {\bfseries 44} (1972) 189--213}.

\bibitem{Bollini:1972ui}
C.~G.~Bollini and J.~J.~Giambiagi, ``{Dimensional Renormalization: The Number of Dimensions as a Regularizing Parameter},'' \href{https://dx.doi.org/10.1007/BF02895558}{Nuovo Cim.\  B {\bfseries 12} (1972) 20--26}.

\bibitem{Breitenlohner:1977hr}
P.~Breitenlohner and D.~Maison, ``{Dimensional Renormalization and the Action Principle},'' \href{https://dx.doi.org/10.1007/BF01609069}{Commun.\  Math.\  Phys.\  {\bfseries 52} (1977) 11--38}.

\bibitem{Belusca-Maito:2020ala}
H.~B{\'e}lusca-Ma{\"\i}to, A.~Ilakovac, M.~Ma{\dj}or-Bo{\v{z}}inovi{\'c}, and D.~St{\"o}ckinger, ``{Dimensional regularization and Breitenlohner-Maison/{\textquoteright}t Hooft-Veltman scheme for $\gamma_5$ applied to chiral YM theories: full one-loop counterterm and RGE structure},'' \href{https://dx.doi.org/10.1007/JHEP08(2020)024}{JHEP {\bfseries 08} (2020) 024} {\ttfamily [\href{https://arxiv.org/abs/2004.14398}{arXiv:2004.14398}]}.

\bibitem{Kinoshita:1958ru}
T.~Kinoshita and A.~Sirlin, ``{Radiative corrections to Fermi interactions},'' \href{https://dx.doi.org/10.1103/PhysRev.113.1652}{Phys.\  Rev.\  {\bfseries 113} (1959) 1652--1660}.

\bibitem{Nir:1989rm}
Y.~Nir, ``{The Mass Ratio $m_c / m_b$ in Semileptonic b-Decays},'' \href{https://dx.doi.org/10.1016/0370-2693(89)91495-0}{Phys.\  Lett.\  B {\bfseries 221} (1989) 184--190}.

\bibitem{vanRitbergen:1998yd}
T.~van Ritbergen and R.~G.~Stuart, ``{Complete two loop quantum electrodynamic contributions to the muon lifetime in the Fermi model},'' \href{https://dx.doi.org/10.1103/PhysRevLett.82.488}{Phys.\  Rev.\  Lett.\  {\bfseries 82} (1999) 488--491} {\ttfamily [\href{https://arxiv.org/abs/hep-ph/9808283}{hep-ph/9808283}]}.

\bibitem{Steinhauser:1999bx}
M.~Steinhauser and T.~Seidensticker, ``{Second order corrections to the muon lifetime and the semileptonic B decay},'' \href{https://dx.doi.org/10.1016/S0370-2693(99)01168-5}{Phys.\  Lett.\  B {\bfseries 467} (1999) 271--278} {\ttfamily [\href{https://arxiv.org/abs/hep-ph/9909436}{hep-ph/9909436}]}.

\bibitem{Ferroglia:1999tg}
A.~Ferroglia, G.~Ossola, and A.~Sirlin, ``{Considerations concerning the radiative corrections to muon decay in the Fermi and standard theories},'' \href{https://dx.doi.org/10.1016/S0550-3213(99)00475-7}{Nucl.\  Phys.\  B {\bfseries 560} (1999) 23--32} {\ttfamily [\href{https://arxiv.org/abs/hep-ph/9905442}{hep-ph/9905442}]}.

\bibitem{Pak:2008qt}
A.~Pak and A.~Czarnecki, ``{Mass effects in muon and semileptonic $b\to c$ decays},'' \href{https://dx.doi.org/10.1103/PhysRevLett.100.241807}{Phys.\  Rev.\  Lett.\  {\bfseries 100} (2008) 241807} {\ttfamily [\href{https://arxiv.org/abs/0803.0960}{arXiv:0803.0960}]}.

\bibitem{Fael:2020tow}
M.~Fael, K.~Sch{\"o}nwald, and M.~Steinhauser, ``{Third order corrections to the semileptonic b{\textrightarrow}c and the muon decays},'' \href{https://dx.doi.org/10.1103/PhysRevD.104.016003}{Phys.\  Rev.\  D {\bfseries 104} (2021) 016003} {\ttfamily [\href{https://arxiv.org/abs/2011.13654}{arXiv:2011.13654}]}.

\bibitem{Czakon:2021ybq}
M.~Czakon, A.~Czarnecki, and M.~Dowling, ``{Three-loop corrections to the muon and heavy quark decay rates},'' \href{https://dx.doi.org/10.1103/PhysRevD.103.L111301}{Phys.\  Rev.\  D {\bfseries 103} (2021) L111301} {\ttfamily [\href{https://arxiv.org/abs/2104.05804}{arXiv:2104.05804}]}.

\bibitem{Marciano:1980pb}
W.~J.~Marciano and A.~Sirlin, ``{Radiative Corrections to Neutrino Induced Neutral Current Phenomena in the $SU(2)_L \times U(1)$ Theory},'' \href{https://dx.doi.org/10.1103/PhysRevD.22.2695}{Phys.\  Rev.\  D {\bfseries 22} (1980) 2695}. [Erratum: Phys.Rev.D 31, 213 (1985)].

\bibitem{Bigi:2023cbv}
D.~Bigi, M.~Bordone, P.~Gambino, U.~Haisch, and A.~Piccione, ``{QED effects in inclusive semi-leptonic B decays},'' \href{https://dx.doi.org/10.1007/JHEP11(2023)163}{JHEP {\bfseries 11} (2023) 163} {\ttfamily [\href{https://arxiv.org/abs/2309.02849}{arXiv:2309.02849}]}. [Erratum: JHEP 03, 078 (2025)].

\bibitem{Marciano:1993sh}
W.~J.~Marciano and A.~Sirlin, ``{Radiative corrections to $\pi_{l2}$ decays},'' \href{https://dx.doi.org/10.1103/PhysRevLett.71.3629}{Phys.\  Rev.\  Lett.\  {\bfseries 71} (1993) 3629--3632}.

\bibitem{Atwood:1989em}
D.~Atwood and W.~J.~Marciano, ``{Radiative Corrections and Semileptonic $B$ Decays},'' \href{https://dx.doi.org/10.1103/PhysRevD.41.1736}{Phys.\  Rev.\  D {\bfseries 41} (1990) 1736}.

\bibitem{Brod:2008ss}
J.~Brod and M.~Gorbahn, ``{Electroweak Corrections to the Charm Quark Contribution to $K^+ \to \pi^+ \nu \bar\nu$},'' \href{https://dx.doi.org/10.1103/PhysRevD.78.034006}{Phys.\  Rev.\  D {\bfseries 78} (2008) 034006} {\ttfamily [\href{https://arxiv.org/abs/0805.4119}{arXiv:0805.4119}]}.

\bibitem{Gorbahn:2022rgl}
M.~Gorbahn, S.~J{\"a}ger, F.~Moretti, and E.~van~der Merwe, ``{Semileptonic weak Hamiltonian to $ \mathcal{O} $({\ensuremath{\alpha}}{\ensuremath{\alpha}}$_{s}$) in momentum-space subtraction schemes},'' \href{https://dx.doi.org/10.1007/JHEP01(2023)159}{JHEP {\bfseries 01} (2023) 159} {\ttfamily [\href{https://arxiv.org/abs/2209.05289}{arXiv:2209.05289}]}.

\bibitem{Ruderman1958}
R.~Gatto and M.~A.~Ruderman, ``{A suggestion on the theory of the $\pi \to e + \nu$ to $\pi \to \mu + \nu$ ratio },'' \href{https://dx.doi.org/10.1007/BF02815260}{Nuovo Cim.\  {\bfseries 8} (1958) 775--777}.

\bibitem{Decker:1994ea}
R.~Decker and M.~Finkemeier, ``{Short and long distance effects in the decay $\tau \to \pi \nu_\tau (\gamma)$},'' \href{https://dx.doi.org/10.1016/0550-3213(95)00597-L}{Nucl.\  Phys.\  B {\bfseries 438} (1995) 17--53} {\ttfamily [\href{https://arxiv.org/abs/hep-ph/9403385}{hep-ph/9403385}]}.

\bibitem{Kinoshita:2001pn}
T.~Kinoshita, ``{Everyone makes mistakes: Including Feynman},'' \href{https://dx.doi.org/10.1088/0954-3899/29/1/302}{J.\  Phys.\  G {\bfseries 29} (2003) 9--22} {\ttfamily [\href{https://arxiv.org/abs/hep-ph/0101197}{hep-ph/0101197}]}.

\bibitem{Desiderio:2020oej}
A.~Desiderio {\em et~al.}, ``{First lattice calculation of radiative leptonic decay rates of pseudoscalar mesons},'' \href{https://dx.doi.org/10.1103/PhysRevD.103.014502}{Phys.\  Rev.\  D {\bfseries 103} (2021) 014502} {\ttfamily [\href{https://arxiv.org/abs/2006.05358}{arXiv:2006.05358}]}.

\bibitem{Giusti:2023pot}
D.~Giusti, C.~F.~Kane, C.~Lehner, S.~Meinel, and A.~Soni, ``{Methods for high-precision determinations of radiative-leptonic decay form factors using lattice QCD},'' \href{https://dx.doi.org/10.1103/PhysRevD.107.074507}{Phys.\  Rev.\  D {\bfseries 107} (2023) 074507} {\ttfamily [\href{https://arxiv.org/abs/2302.01298}{arXiv:2302.01298}]}.

\bibitem{Frezzotti:2023ygt}
R.~Frezzotti, {\em et al.}, ``{Lattice calculation of the Ds meson radiative form factors over the full kinematical range},'' \href{https://dx.doi.org/10.1103/PhysRevD.108.074505}{Phys.\  Rev.\  D {\bfseries 108} (2023) 074505} {\ttfamily [\href{https://arxiv.org/abs/2306.05904}{arXiv:2306.05904}]}.

\bibitem{BESIII:2021anh}
{\bfseries BESIII} Collaboration, ``{Measurement of the absolute branching fractions for purely leptonic $D_s^+$ decays},'' \href{https://dx.doi.org/10.1103/PhysRevD.104.052009}{Phys.\  Rev.\  D {\bfseries 104} (2021) 052009} {\ttfamily [\href{https://arxiv.org/abs/2102.11734}{arXiv:2102.11734}]}.

\bibitem{BESIII:2023cym}
{\bfseries BESIII} Collaboration, ``{Improved measurement of the branching fraction of $D_s^+ \to {\ensuremath{\mu}}^+{\ensuremath{\nu}}_{\ensuremath{\mu}}$},'' \href{https://dx.doi.org/10.1103/PhysRevD.108.112001}{Phys.\  Rev.\  D {\bfseries 108} (2023) 112001} {\ttfamily [\href{https://arxiv.org/abs/2307.14585}{arXiv:2307.14585}]}.

\bibitem{BESIII:2024dvk}
{\bfseries BESIII} Collaboration, ``{Measurement of the branching fraction of $D_s^+\to {\ensuremath{\ell}}^+{\ensuremath{\nu}}_{\ensuremath{\ell}}$ via $e^+e^- \to D_s^{*+}D_s^{*-}$},'' \href{https://dx.doi.org/10.1103/PhysRevD.110.052002}{Phys.\  Rev.\  D {\bfseries 110} (2024) 052002} {\ttfamily [\href{https://arxiv.org/abs/2407.11727}{arXiv:2407.11727}]}.

\bibitem{BESIII:2021bdp}
{\bfseries BESIII} Collaboration, ``{Measurement of the Absolute Branching Fraction of $D_s^+ \to \tau^+ \nu_{\tau}$ via $\tau^+ \to e^+ \nu_e \bar{\nu}_{\tau}$},'' \href{https://dx.doi.org/10.1103/PhysRevLett.127.171801}{Phys.\  Rev.\  Lett.\  {\bfseries 127} (2021) 171801} {\ttfamily [\href{https://arxiv.org/abs/2106.02218}{arXiv:2106.02218}]}.

\bibitem{BESIII:2021wwd}
{\bfseries BESIII} Collaboration, ``{Measurement of the branching fraction of leptonic decay $D_s^+\to\tau^+\nu_\tau$ via $\tau^+\to\pi^+\pi^0\bar \nu_\tau$},'' \href{https://dx.doi.org/10.1103/PhysRevD.104.032001}{Phys.\  Rev.\  D {\bfseries 104} (2021) 032001} {\ttfamily [\href{https://arxiv.org/abs/2105.07178}{arXiv:2105.07178}]}.

\bibitem{BESIII:2023ukh}
{\bfseries BESIII} Collaboration, ``{Measurement of the branching fraction of $D_s^+ \to \tau^+ \nu_\tau$ via $\tau^+ \to \mu^+ \nu_\mu \bar\nu_\tau$},'' \href{https://dx.doi.org/10.1007/JHEP09(2023)124}{JHEP {\bfseries 09} (2023) 124} {\ttfamily [\href{https://arxiv.org/abs/2303.12468}{arXiv:2303.12468}]}.

\bibitem{BESIII:2023fhe}
{\bfseries BESIII} Collaboration, ``{Updated measurement of the branching fraction of $D_s^+\to {\ensuremath{\tau}}^+{\ensuremath{\nu}}_{\ensuremath{\tau}}$ via ${\ensuremath{\tau}}^+\to {\ensuremath{\pi}}^+ \bar{\nu}_\tau$},'' \href{https://dx.doi.org/10.1103/PhysRevD.108.092014}{Phys.\  Rev.\  D {\bfseries 108} (2023) 092014} {\ttfamily [\href{https://arxiv.org/abs/2303.12600}{arXiv:2303.12600}]}.

\bibitem{Kinoshita:1962ur}
T.~Kinoshita, ``{Mass singularities of Feynman amplitudes},'' \href{https://dx.doi.org/10.1063/1.1724268}{J.\  Math.\  Phys.\  {\bfseries 3} (1962) 650--677}.

\bibitem{Lee:1964is}
G.~Feinberg, ed., ``{Degenerate Systems and Mass Singularities},'' \href{https://dx.doi.org/10.1103/PhysRev.133.B1549}{Phys.\  Rev.\  {\bfseries 133} (1964) B1549--B1562}.

\bibitem{Bryman:1982et}
D.~A.~Bryman, P.~Depommier, and C.~Leroy, ``{$\pi \to e \nu,~\pi \to e \nu \gamma$ decays and related processes},'' \href{https://dx.doi.org/10.1016/0370-1573(82)90162-4}{Phys.\  Rept.\  {\bfseries 88} (1982) 151--205}.

\bibitem{Yennie:1961ad}
D.~R.~Yennie, S.~C.~Frautschi, and H.~Suura, ``{The infrared divergence phenomena and high-energy processes},'' \href{https://dx.doi.org/10.1016/0003-4916(61)90151-8}{Annals Phys.\  {\bfseries 13} (1961) 379--452}.

\bibitem{Weinberg:1965nx}
S.~Weinberg, ``{Infrared photons and gravitons},'' \href{https://dx.doi.org/10.1103/PhysRev.140.B516}{Phys.\  Rev.\  {\bfseries 140} (1965) B516--B524}.

\bibitem{Isidori:2007zt}
G.~Isidori, ``{Soft-photon corrections in multi-body meson decays},'' \href{https://dx.doi.org/10.1140/epjc/s10052-007-0487-0}{Eur.\  Phys.\  J.\  C {\bfseries 53} (2008) 567--571} {\ttfamily [\href{https://arxiv.org/abs/0709.2439}{arXiv:0709.2439}]}.

\bibitem{Kinoshita:1959ha}
T.~Kinoshita, ``{Radiative corrections to $\pi - e$ decay},'' \href{https://dx.doi.org/10.1103/PhysRevLett.2.477}{Phys.\  Rev.\  Lett.\  {\bfseries 2} (1959) 477}.

\bibitem{Rowe:2024pfs}
M.~Rowe, \href{https://dx.doi.org/10.7488/era/5089}{{\em {Structure-dependent quantum electrodynamics in heavy meson physics}}}.
\newblock PhD thesis, Edinburgh U., 2024.

\bibitem{Barberio:1990ms}
E.~Barberio, B.~van Eijk, and Z.~Was, ``{PHOTOS: A Universal Monte Carlo for QED radiative corrections in decays},'' \href{https://dx.doi.org/10.1016/0010-4655(91)90012-A}{Comput.\  Phys.\  Commun.\  {\bfseries 66} (1991) 115--128}.

\bibitem{Barberio:1993qi}
E.~Barberio and Z.~Was, ``{PHOTOS: A Universal Monte Carlo for QED radiative corrections. Version 2.0},'' \href{https://dx.doi.org/10.1016/0010-4655(94)90074-4}{Comput.\  Phys.\  Commun.\  {\bfseries 79} (1994) 291--308}.

\bibitem{Golonka:2005pn}
P.~Golonka and Z.~Was, ``{PHOTOS Monte Carlo: A Precision tool for QED corrections in $Z$ and $W$ decays},'' \href{https://dx.doi.org/10.1140/epjc/s2005-02396-4}{Eur.\  Phys.\  J.\  C {\bfseries 45} (2006) 97--107} {\ttfamily [\href{https://arxiv.org/abs/hep-ph/0506026}{hep-ph/0506026}]}.

\bibitem{Davidson:2010ew}
N.~Davidson, T.~Przedzinski, and Z.~Was, ``{PHOTOS interface in C++: Technical and Physics Documentation},'' \href{https://dx.doi.org/10.1016/j.cpc.2015.09.013}{Comput.\  Phys.\  Commun.\  {\bfseries 199} (2016) 86--101} {\ttfamily [\href{https://arxiv.org/abs/1011.0937}{arXiv:1011.0937}]}.

\bibitem{Parrott:2022rgu}
{\bfseries HPQCD} Collaboration, ``{B{\textrightarrow}K and D{\textrightarrow}K form factors from fully relativistic lattice QCD},'' \href{https://dx.doi.org/10.1103/PhysRevD.107.014510}{Phys.\  Rev.\  D {\bfseries 107} (2023) 014510} {\ttfamily [\href{https://arxiv.org/abs/2207.12468}{arXiv:2207.12468}]}.

\bibitem{FermilabLattice:2022gku}
{\bfseries Fermilab Lattice, MILC} Collaboration, ``{D-meson semileptonic decays to pseudoscalars from four-flavor lattice QCD},'' \href{https://dx.doi.org/10.1103/PhysRevD.107.094516}{Phys.\  Rev.\  D {\bfseries 107} (2023) 094516} {\ttfamily [\href{https://arxiv.org/abs/2212.12648}{arXiv:2212.12648}]}.

\bibitem{Lubicz:2017syv}
{\bfseries ETM} Collaboration, ``{Scalar and vector form factors of $D \to \pi(K) \ell \nu$ decays with $N_f=2+1+1$ twisted fermions},'' \href{https://dx.doi.org/10.1103/PhysRevD.96.054514}{Phys.\  Rev.\  D {\bfseries 96} (2017) 054514} {\ttfamily [\href{https://arxiv.org/abs/1706.03017}{arXiv:1706.03017}]}. [Erratum: Phys.Rev.D 99, 099902 (2019), Erratum: Phys.Rev.D 100, 079901 (2019)].

\bibitem{Lubicz:2018rfs}
{\bfseries ETM} Collaboration, ``{Tensor form factor of $D \to \pi(K) \ell \nu$ and $D \to \pi(K) \ell \ell$ decays with $N_f=2+1+1$ twisted-mass fermions},'' \href{https://dx.doi.org/10.1103/PhysRevD.98.014516}{Phys.\  Rev.\  D {\bfseries 98} (2018) 014516} {\ttfamily [\href{https://arxiv.org/abs/1803.04807}{arXiv:1803.04807}]}.

\bibitem{Rowe:2024jml}
M.~Rowe and R.~Zwicky, ``{Structure-dependent QED in $ {B}^{-}\to {\ell}^{-}\overline{\nu}\left(\gamma \right) $},'' \href{https://dx.doi.org/10.1007/JHEP07(2024)249}{JHEP {\bfseries 07} (2024) 249} {\ttfamily [\href{https://arxiv.org/abs/2404.07648}{arXiv:2404.07648}]}.

\bibitem{Isidori:2020acz}
G.~Isidori, S.~Nabeebaccus, and R.~Zwicky, ``{QED corrections in $ \overline{B}\to \overline{K}{\mathrm{\ell}}^{+}{\mathrm{\ell}}^{-} $ at the double-differential level},'' \href{https://dx.doi.org/10.1007/JHEP12(2020)104}{JHEP {\bfseries 12} (2020) 104} {\ttfamily [\href{https://arxiv.org/abs/2009.00929}{arXiv:2009.00929}]}.

\end{thebibliography}\endgroup

\end{document}